\documentclass[10pt,letterpaper]{article}
\usepackage[top=0.85in,left=2.75in,footskip=0.75in]{geometry}

\usepackage{amsmath,amssymb}

\usepackage{changepage}

\usepackage[utf8x]{inputenc}

\usepackage{textcomp,marvosym}

\usepackage{cite}

\usepackage{nameref,hyperref}

\usepackage[right]{lineno}

\usepackage{microtype}
\DisableLigatures[f]{encoding = *, family = * }

\usepackage[table]{xcolor}

\usepackage{array}

\newcolumntype{+}{!{\vrule width 2pt}}

\newlength\savedwidth

\usepackage[aboveskip=1pt,labelfont=bf,labelsep=period,justification=raggedright,singlelinecheck=off]{caption}
\renewcommand{\figurename}{Fig}
\newcommand{\figurenames}{Figs}

\makeatletter
\renewcommand{\@biblabel}[1]{\quad#1.}
\makeatother

\usepackage{lastpage,fancyhdr,graphicx}
\usepackage{epstopdf}
\fancyheadoffset[L]{2.25in}
\fancyfootoffset[L]{2.25in}
\newcommand{\muv}{\boldsymbol{\mu}}

\usepackage{comment}
\graphicspath{{imgs/}}

\begin{document}
\vspace*{0.2in}

\begin{flushleft}
{\Large
\textbf\newline{Deep learning-based reduced order models in cardiac electrophysiology} 
}
\newline
\\
Stefania Fresca\textsuperscript{1},
Andrea Manzoni\textsuperscript{1*},
Luca Ded\'e\textsuperscript{1},
Alfio Quarteroni\textsuperscript{1,2},
\\
\bigskip
\textbf{1} \footnotesize MOX - Dipartimento di Matematica, Politecnico di Milano, Milano, Italy
\\
\textbf{2} \footnotesize Mathematics Institute, \'Ecole Polytechnique F\'ed\'erale de Lausanne, Lausanne, Switzerland
\\
\bigskip

%
%





* andrea1.manzoni@polimi.it

\end{flushleft}
\section*{Abstract}

Predicting the electrical behavior of the heart, from the cellular scale to the tissue level, relies on the formulation and numerical approximation of coupled nonlinear dynamical systems. These systems describe the cardiac action potential, that is the polarization/depolarization cycle occurring at every heart beat that models the time evolution of the electrical potential across the cell membrane, as well as a set of ionic variables. Multiple solutions of these systems, corresponding to different model inputs, are required to evaluate outputs of clinical interest, such as activation maps and action potential duration. More importantly, these models feature coherent structures that propagate over time, such as wavefronts. 
These systems can hardly be reduced to lower dimensional problems by conventional reduced order models (ROMs) such as, e.g., the reduced basis (RB) method. This is primarily due to the low regularity of the solution manifold (with respect to the problem parameters) as well as to the nonlinear nature of the input-output maps that we intend to reconstruct numerically. To overcome this difficulty, in this paper
we propose a new, nonlinear approach which exploits deep learning (DL) algorithms to obtain accurate and efficient ROMs, whose dimensionality matches the number of system parameters. Our DL approach combines deep feedforward neural networks (NNs) and convolutional autoencoders (AEs). 
We show that the proposed DL-ROM framework can efficiently provide solutions to parametrized electrophysiology problems, thus enabling multi-scenario analysis in pathological cases. We investigate three challenging test cases in cardiac electrophysiology and prove that DL-ROM outperforms classical projection-based ROMs. 


\section*{Introduction}

The electrical activation of the heart is the main responsible of its contraction, is the result of two processes: at the microscopic scale, the generation of ionic currents through the cellular membrane producing a local action potential; and at the macroscopic scale, the propagation of the action potential from cell to cell   in the form of a transmembrane potential \cite{quarteroni_manzoni_vergara_2017, quarteroni_dede'_manzoni_vergara_2019, franzone2014mathematical}. This latter process can be described by means of partial differential equations (PDEs), suitably coupled with systems of ordinary differential equations (ODEs) modeling the ionic currents in the cells. 

Solving this system using a high-fidelity, full order model (FOM) such as, e.g., the finite element (FE) method, is computationally demanding. Indeed, the propagation of the electrical signal is characterized by the fast dynamics of very steep fronts, thus requiring very fine space and time discretizations. 
\cite{sundnes2007computing,colli2004parallel,franzone2014mathematical}.
Using a FOM may quickly become unaffordable if such a coupled system must be solved for several values of parameters representing either functional or geometric data such as, e.g., material properties, initial and boundary conditions, or the shape of the domain. Multi-query analysis is relevant in a variety of situations: when analysing multiple scenarios, when dealing with sensitivity analysis and uncertainty quantification (UQ) problems in order to account for inter-subject variability \cite{Mirams_2016,Johnstone201649,Hurtado_2017}, for parameter estimation or data assimilation, in which some unknown (or unaccessible) quantities characterizing the mathematical model must be inferred from a set of measurements \cite{dhamala2018quantifying,quaglino2018fast,johnston2018quantifying,pathmanathan2019comprehensive,levrero2020sensitivity}.

Conventional projection-based reduced order models (ROMs) built, e.g., through the reduced basis (RB) method \cite{quarteroni2016reduced}, yields inefficient ROMs when dealing with nonlinear time-dependent parametrized PDE-ODE system  as the one arising from cardiac electrophysiology. 
The  {three major computational bottlenecks} shown by such kind of ROMs for cardiac electrophysiology are due the fact that: 
\begin{itemize}
\item[-] the linear superimposition of modes, on which they are based, would cause the dimension of the ROM to be excessively large to guarantee an acceptable accuracy;
\item[-] evaluating the ROM requires the solution of a dynamical system, which might be unstable unless the size of time step $\Delta t$ is very small;
\item[-] the ROM must also account for the dynamics of the gating variables, even when aiming at computing just the electrical potential. This fact entails an extremely intrusive and costly hyper-reduction stage   
to reduce the solution of the ODE system to a few, selected mesh nodes  \cite{pagani2018numerical}.
\end{itemize} 

To overcome the limitations of projection-based ROMs, we propose a new, non-intrusive ROM technique based on deep learning (DL) algorithms, which we refer to as DL-ROM. Combining in a suitable way a convolutional autoencoder (AE) and a deep feedforward neural network (DFNN), the DL-ROM technique enables the construction of an efficient ROM, whose dimension is as close as possible to the number of parameters upon which the solution of the differential problem depends. 
A preliminary numerical assessment of our DL-ROM technique has already been presented in \cite{fresca2020comprehensive}, albeit on simpler -- yet challenging -- test cases. 

The proposed DL-ROM technique is a combination of a data-driven with a physics based model approach. Indeed, it exploits snapshots taken from a set of FOM solutions (for selected parameter values and time instances) and deep neural network architectures to learn, in a non-intrusive way, both {\em (i)} the nonlinear trial manifold where the ROM solution is sought, and  {\em (ii)} the nonlinear reduced dynamics. In a linear ROM built, e.g., thorugh proper orthogonal decomposition (POD), the former quantity is nothing but a set of basis functions, while the latter task corresponds to the projection stage in the subspace spanned by these basis functions. Here, our goal is to show that DL-ROM can be effectively used to handle parametrized problems in cardiac electrophysiology, accounting for both physiological and pathological conditions, in order to provide fast and accurate solutions. The proposed DL-ROM is  computationally efficient during the testing stage, that is for any new scenario unseen during the training stage. This is particularly useful in view of the evaluation of patient-specific features to enable the integration of computational methods in  current clinical platforms.
 

DL techniques for parametrized PDEs have previously been proposed in other contexts. In  \cite{guo2018reduced, guo2019data, hestaven2018non-intrusive, san2018neural} feedforward neural networks have been employed to model the reduced dynamics in a less intrusive way, that is, avoiding the costs entailed by projection-based ROMs, but still relying on a linear trial manifold built, e.g., through POD. In  \cite{kani2017dr-rnn, mohan2018adeep, wan2018data} the construction of  ROMs for nonlinear, time-dependent problems has been replaced by the evaluation of ANN-based regression models. In \cite{gonzalez2018deep, carlberg2018model} the reduced trial manifold where the approximation is sought has been modeled through ANNs thus avoiding  the linear superimposition of POD modes, on a minimum residual formulation to derive the ROM \cite{carlberg2018model}, or without considering an explicit parameter dependence in the differential problem that is considered \cite{gonzalez2018deep}. In all these works, coupled problems have never been considered. Moreover, very often DL techniques have been exploited to address problems which require only a moderate dimension of projection-based ROMs. We demonstrate that our DL-ROM provides accurate results by constructing ROMs with extremely low-dimension in prototypical test cases. These tests exhibit all the relevant physical features which make the numerical approximation of parametrized problems in cardiac electrophysiology a challenging task.

\section*{Materials and methods}
\subsection*{Cardiac electrophysiology}

Muscle contraction and relaxation drive the pump function of the heart. In particular, tissue contraction is triggered by electrical signals self-generated in the heart and propagated through the myocardium thanks to the excitability of the cardiac cells, the cardiomyocites \cite{franzone2014mathematical, klabunde2011cardiovascular}. When suitably stimulated, cardiomyocites produce a variation of the potential across the cellular membrane, called {\em transmembrane potential}. Its evolution in time is usually referred to as {\em action potential}, involving a polarization and a depolarization in the early stage of every heart beat. The action potential is generated by several ion channels (e.g., calcium, sodium, potassium) that open and close, and by the resulting ionic currents crossing the membrane. For instance, coupling the so-called monodomain model for the transmembrane potential $u = u({\bf x}, t)$ with a phenomenological model for the ionic currents -- involving a single gating variable $w = w({\bf x},t)$ -- in a domain $\Omega$ representing, e.g., a portion of the myocardium, results in the following  nonlinear time-dependent system  \vspace{-0.1cm}
\begin{equation}
\label{eq:monodomain}
\begin{cases}
\displaystyle \frac{\partial u}{\partial t} - \textnormal{div}({\bf D} \nabla u) +I_{ion}(u, w) = I_{app}(\mathbf{x}, t) \quad & (\mathbf{x}, t) \in \Omega \times (0,T), \vspace{0.1cm} \\ 
\displaystyle \frac{\partial w}{\partial t} + g(u,w)=0 \quad & (\mathbf{x}, t) \in \Omega \times (0,T), \vspace{0.1cm}\\
\displaystyle \nabla u \cdot \mathbf{n} = 0 \quad & (\mathbf{x}, t) \in \partial \Omega \times (0,T),\\
u(\mathbf{x},0)=0, \; w(\mathbf{x},0)= 0 \quad & \mathbf{x} \in \Omega. \vspace{-0.1cm}
\end{cases}
\end{equation}
Here $t$ denotes a rescaled time\footnote{Dimensional times and potential  \cite{AlievPanfilov} are given by $\tilde{t} [ms] = 12.9 t$ and $\tilde{u} [mV] = 100 u - 80$. The transmembrane potential ranges  from the resting state of $-80$ mV to the excited state of $+20$ mV.}, ${\bf n}$ denotes the outward directed unit vector normal to the boundary $\partial \Omega$ of  $\Omega$, whereas $I_{app}$ is an applied current representing, e.g., the initial activation of the tissue. The nonlinear diffusion-reaction equation for $u$ is two-ways coupled with the ODE system, which must be in principle solved at any point ${\bf x} \in \Omega$; indeed, the reaction term $I_{ion}$ and the function $g$ depend on both $u$ and $w$. The most common choices for the two functions $I_{ion}$ and $g$ in order to efficiently reproduce the action-potential are, e.g., the FitzHugh-Nagumo \cite{fitzhugh1961impulses,nagumo1962active}, the Aliev-Panfilov
\cite{AlievPanfilov,NashPanfilov} or the Mitchell and Schaeffer models \cite{mitchell2003two}. The diffusivity tensor ${\bf D}$ usually depends on the fibers-sheet structure of the tissue, affecting directional conduction velocities and directions. In particular, by assuming an axisymmetric distribution of the fibers, the conductivity tensor takes the form  
\begin{equation}
\label{eq:conductivity_tensor}
\bf{D}(\mathbf{x}) = \sigma_t I + (\sigma_l - \sigma_t) \mathbf{f}_0 \otimes \mathbf{f}_0, 
\end{equation}
where $\sigma_l$ and  $\sigma_t$ are   the conductivities in the fibers and the  transversal directions.

When a simple phenomenological ionic model is considered, such as the  FitzHugh-Nagumo or the Aliev-Panfilov (A-P) model, the ionic current takes the form of a cubic nonlinear function of $u$ and a single (dimensionless) gating variable plays the role of a recovery function, allowing to model refractariness of cells. 
In this paper, we focus on the Aliev-Panfilov model, which consists in taking
\begin{equation}
\label{eq:aliev_panfilov}
\begin{aligned}
& I_{ion}(u,w) = Ku(u - a)(u - 1) + uw, \\
& g(u, w) = \displaystyle \Big( \epsilon_0 + \frac{c_1w}{c_2+u} \Big)(- w - Ku(u - b -1)).
\end{aligned}
\end{equation}
The parameters $K$, $a$, $b$, $\varepsilon_0$, $c_1$, $c_2$ are related to the cell. 
Here $a$ represents an oscillation threshold, whereas the  weighting factor $\varepsilon_0 +\frac{ c_1 w}{c_2 + u}$ 
was introduced in \cite{AlievPanfilov} to tune the restitution curve to experimental observations by adjusting the parameters $c_1$ and $c_2$; see, e.g., \cite{CLAYTON201122,quarteroni_manzoni_vergara_2017, quarteroni_dede'_manzoni_vergara_2019,franzone2014mathematical} for a detailed review. In the remaining part of the paper, we denote by $\muv \in \mathcal{P} \subset {\mathbb{R}}^{n_{\boldsymbol{\mu}}}$ a parameter vector listing all the ${n}_{\boldsymbol{\mu}}$ input parameters characterizing physical (and, possibly, geometrical) properties we might be interested to vary; $\mathcal{P}$ is a subset of ${\mathbb{R}}^{n_{\boldsymbol{\mu}}}$, denoting the parameter space. Relevant physical situations are those in which input parameters affect the diffusivity matrix $\mathbf{D}$ (through the conduction velocities) and the applied current $I_{app}$; previous analyses focused instead on the  gating variable dynamics   (through   $g$) and the ionic current $I_{ion}$, see \cite{pagani2018numerical}.

\subsection*{Projection-based ROMs}
From an algebraic standpoint, the spatial discretization of system \eqref{eq:monodomain} through the Galerkin-finite element (FE) approximation \cite{quarteroni1994numerical} yields the following nonlinear dynamical system for ${\bf u} = {\bf u}(t; \muv)$, ${\bf w} = {\bf w}(t; \muv)$, representing our full order model (FOM): \vspace{-0.1cm}
\begin{equation}
\label{eq:FOM}
\left\{
\begin{array}{ll}
\displaystyle \mathbf{M}(\muv) 	\frac{\partial \mathbf{{u}}}{\partial t}  =    \mathbf{A}(\muv) \mathbf{u} + \mathbf{I}_{ion}(t, \mathbf{u} , \mathbf{w}; \boldsymbol{\mu}) + {\bf I}_{app}(t; \muv), & \ \  t \in (0, T),   \\
\displaystyle \frac{\partial \mathbf{{w}}}{\partial t} (t;\boldsymbol{\mu}) = \mathbf{g}(t, \mathbf{u}, \mathbf{w}; \boldsymbol{\mu}), & \ \  t \in (0, T), \smallskip \\
\mathbf{u}(0) =\mathbf{u}_0,  \qquad 
\mathbf{w}(0) =\mathbf{w}_0. 
\end{array} \vspace{-0.1cm}
\right.
\end{equation}
Here $\mathbf{A}(\muv) \in {\mathbb{R}}^{N \times N}$  is a matrix arising from the diffusion operator (thus including the conductivity tensor  $\mathbf{D}(\muv)=\mathbf{D}(\mathbf{x};\muv)$, which can vary within the myocardium due to fiber orientation and conditions, such as  the possible presence of ischemic regions);   $\mathbf{M}(\muv) \in {\mathbb{R}}^{N \times N}$ is the mass matrix; $\mathbf{I}_{ion}, \mathbf{g} \in {\mathbb{R}}^{N}$ are vectors arising from the nonlinear terms; ${\bf I}_{app} \in {\mathbb{R}}^{N}$ is a vector collecting the applied currents; finally, $\mathbf{u}_0, \mathbf{w}_0 \in {\mathbb{R}}^{N}$ are the initial data, possibly depending on $\muv$. The dimension $N$ is related to the dimension of the FE space and, ultimately, depends on the size $h>0$ of the computational mesh used to discretize the domain $\Omega$. Note that the system of ODEs arises from the collocation of the ODE \eqref{eq:monodomain}${}_2$ at the nodes used for the numerical integration. 

The intrinsic dimension of the solution manifold 
\begin{equation}
\mathcal{S} = \{ \mathbf{u}(t ; \boldsymbol{\mu} ) \; | \; t \in [0, T) \; \textnormal{and} \; \boldsymbol{\mu} \in \mathcal{P} \subset {\mathbb{R}}^{n_{\mu}} \} \subset {\mathbb{R}}^{N}, 
\label{eq:solution_manifold}
\end{equation}
obtained by solving (\ref{eq:FOM}) when $(t ; \boldsymbol{\mu} )$ varies in $[0, T) \times \mathcal{P} $, is usually much smaller than $N$ and, under suitable conditions, is at most $n_{\boldsymbol{\mu}} + 1  \ll N$, where $n_{\boldsymbol \mu}$ is the number of parameters -- in this respect, the time independent variable plays the role of a parameter. For this reason, ROMs attempt at  approximating $\mathcal{S}$ by introducing a suitable trial manifold of lower dimension. The most popular approach is {\em proper orthogonal decomposition} (POD), which exploits a  \textit{linear trial manifold} built through the singular value decomposition of a matrix ${\bf S} \in {\mathbb{R}}^{N \times N_s}$ collecting a set of FOM snapshots
\begin{equation*}
{\bf S} = \left[ \mathbf{u}({t}^1; \boldsymbol {\mu}_1) \; | \; \ldots \; | \; \mathbf{u}({t}^{N_t}; \boldsymbol {\mu}_1) \; | \; \ldots \; | \;
\mathbf{u}({t}^1 ; \boldsymbol {\mu}_{N_{train}}) \; | \; \ldots \; | \; \mathbf{u}({t}^{N_t} ; \boldsymbol {\mu}_{N_{train}}) \right]; \vspace{0.1cm}
\end{equation*}
this is a set of solutions obtained for $N_{train}$ selected input parameters at (a subset, possibly, of) the time instants $\{{t}^k\}_{k=1}^{N_t}$ in which $(0,T)$ is partitioned for the sake of time discretization.  The most common choice is to set ${t}^k = k \Delta t$ where $\Delta t = T/(N_t - 1)$.  

When using a projection-based ROM, the approximation of $\mathbf{u}(t; \muv)$ is sought as a linear superimposition of modes, under the form
\begin{equation} \label{eq:RB1}
\mathbf{u}(t;\muv) \approx {\bf V}  \mathbf{u}_{n}(t;\muv),
\end{equation}
thus yielding a linear ROM, in which the columns of the matrix ${\bf V} = [\boldsymbol{\zeta}_1, \ldots, \boldsymbol{\zeta}_n] \in {\mathbb{R}}^{N \times n}$ form an orthonormal basis of a space $V_n$, an $n$-dimensional subspace of ${\mathbb{R}}^{N}$. In the case of POD, $V_n$ provides the best $n$-rank approximation of ${\bf S}$ in the Frobenius norm, that is, $\boldsymbol{\zeta}_1, \ldots, \boldsymbol{\zeta}_n$ are the first $n$ (left) singular vectors of ${\bf S}$ corresponding to the $n$ largest singular values $ \sigma_1,\ldots,  \sigma_n$ of ${\bf S}$,  {such that  the projection error is smaller than a desired tolerance $\varepsilon_{POD}$. To meet this requirement, it is sufficient to choose $n$ as the smallest integer such that
\[
\frac{	\sum_{i=1}^{N} \sigma_i^2 }{	\sum_{i=1}^{N_s} \sigma_i^2} >1- {\varepsilon}_{POD}^2,
\]
i.e., the energy retained by the last $N_s - n$ POD modes is equal or smaller than  ${\varepsilon}_{POD}^2$.}

The approximation of $\mathbf{w}$ is given instead by its restriction 
\begin{equation*}
\mathbf{w}(t;\muv) \approx {\bf P}  \mathbf{w}_{m}(t;\muv),
\end{equation*}
to a (possibly, small)  subset of $m$ degrees of freedom, where $m \ll n$, at which the nonlinear term ${\bf I}_{ion}$ is interpolated exploiting  a problem-dependent basis, spanned by the columns of a matrix  $\boldsymbol{\Phi}  \in {\mathbb{R}}^{N \times m}$, which is built according to a suitable {\em hyper-reduction} strategy; see, e.g., \cite{pagani2018numerical} for further details.  Here    ${\bf P}  = [{\bf e}_1, \ldots, {\bf e}_m] \in  {\mathbb{R}}^{N  \times m}$ denotes a matrix formed by the columns of the $N \times N$ identity matrix corresponding to the $m$ selected degrees of freedom.

A Galerkin-POD ROM for system \eqref{eq:monodomain} is then obtained by {\em (i)} first, substituting equation \eqref{eq:RB1} into equation \ref{eq:FOM} and projecting it onto $V_n$; then, {\em (ii)} solving the system of ODEs at $m$ selected degrees of freedom, thus yielding the following  nonlinear dynamical system for ${\bf u}_n = {\bf u}_n(t; \muv)$ and the selected components ${\mathbf{P}}^T \mathbf{w} = {\mathbf{P}}^T \mathbf{w} (t; \muv)$ of  ${\bf w}$:
\begin{equation} \label{eq:ROM-RB}
  \begin{cases}
     \displaystyle    {\mathbf{V}}^T \mathbf{M}(\muv) \mathbf{V} \frac{\partial \mathbf{u}_n}{\partial t}  +  {\mathbf{V}}^T \mathbf{A} (\muv) {\mathbf{V}}^T  \mathbf{u}_n    
     \smallskip   \\
     \hspace{0.5cm}  +  {\mathbf{V}}^T \boldsymbol{\Phi} {({\mathbf{P}}^T \boldsymbol{\Phi})}^{-1} \mathbf{I}_{ion}( t, {\mathbf{P}}^T \mathbf{V} \mathbf{u}_n , {\mathbf{P}}^T \mathbf{w}  ; \muv )-  {\bf V}^T \mathbf{I}_{app} (t; \muv) = {\bf 0},  &   t \in (0,T),     
     \smallskip   \\
    \displaystyle  {\mathbf{P}}^T \frac{\partial \mathbf{w}}{\partial t}  +  \mathbf{g}(t, {\mathbf{P}}^T \mathbf{V}\mathbf{u}_n, {\mathbf{P}}^T \mathbf{w}  ; \muv ) = {\bf 0}, &   t \in (0,T)  ,  \smallskip \\
     \mathbf{u}_n(0)  = {\bf V}^T \mathbf{u}_0, \qquad 
    {\mathbf{P}}^T \mathbf{w}(0)  = {\bf P}^T \mathbf{w}_0.
   \end{cases}
\end{equation}
 
 This strategy is the essence of the reduced basis (RB) method for nonlinear time-dependent parametrized PDEs. However, using \eqref{eq:ROM-RB} as an approximation to \eqref{eq:FOM} is known to suffer from several problems. First of all, an extensive hyper-reduction stage (exploiting, e.g., the discrete empirical interpolation method (DEIM)) must be performed in order to be able to evaluate any $\muv$- or ${\bf u}$-dependent quantities appearing in \eqref{eq:ROM-RB}, that is, without relying on $N$-dimensional arrays. Moreover, whenever the solution of the differential problem  features coherent structures that propagate over time, such as steep wavefronts,  the dimension $n$ of the projection-based ROM \eqref{eq:ROM-RB} might easily become very large, due to the basic linearity assumption, by which the solution is given by a linear superimposition of POD modes, thus severely degrading the computational efficiency of the ROM. A possible way to overcome this bottleneck is to rely on local reduced bases, built through POD after the set of snapshots has been split into $N_c >1$ clusters, according to suitable clustering (or unsupervised learning) algorithms \cite{pagani2018numerical}. 
\subsection*{Deep learning-based reduced order modeling (DL-ROM)}

To overcome the limitations of linear ROMs, we consider  a new, nonlinear ROM technique based on deep learning models. First introduced in \cite{fresca2020comprehensive} and assessed on one-dimensional benchmark problems, the DL-ROM technique aims at learning both the nonlinear trial manifold (corresponding to the matrix ${\bf V}$  in the case of a  linear ROM) in which we seek the solution to the parametrized system \eqref{eq:monodomain}  and the nonlinear reduced dynamics (corresponding to the projection stage in a linear ROM). This method is not intrusive; it relies on DL algorithms trained on a set of FOM solutions obtained for different parameter values. 


We denote by $N_{train}$ and $N_{test}$ the number of training and testing parameter instances, respectively; the ROM dimension is again denoted by $n  \ll N$. 
In order to describe the system dynamics on a suitable reduced nonlinear trial manifold (a task which we refer to as {\em reduced dynamics learning}), the intrinsic coordinates of the ROM approximation are defined as 
\begin{equation}
\label{eq:h}
{\mathbf{u}}_n(t; \boldsymbol \mu, \boldsymbol{\theta}_{DF}) = {\boldsymbol{\phi}}_n^{DF}(t; \boldsymbol \mu, \boldsymbol{\theta}_{DF}), 
\end{equation}
where ${\boldsymbol{\phi}}_n^{DF}(\cdot; \cdot, \boldsymbol{\theta}_{DF}) : {\mathbb{R}}^{(n_{\mu} +1)} \rightarrow {\mathbb{R}}^n$ is a deep feedforward neural network (DFNN), consisting in the subsequent composition of a nonlinear activation function,  applied to a linear transformation of the input, multiple times \cite{goodfellow2016deep}.  Here $\boldsymbol{\theta}_{DF}$ denotes the vector of {parameters} of the DFNN, collecting all the corresponding weights and biases of each layer of the DFNN.

Regarding instead the description of the reduced nonlinear trial manifold, approximating the solution one, $\tilde{\mathcal{S}} \approx \mathcal{S}$ (a task which we refer to as  {\em reduced trial manifold learning}) we employ the \textit{decoder function of a convolutional autoencoder}\footnote{The AE is a particular type of neural network aiming at learning the identity function
\begin{equation}
\mathbf{f}^{AE} (\cdot; \boldsymbol{\theta}_E, \boldsymbol{\theta}_D) : \mathbf{x} \mapsto \mathbf{\tilde{x}} \quad \textnormal{with} \quad \mathbf{\tilde{x}} \simeq \mathbf{x}.
\end{equation}.
It is composed by two main parts:
\begin{itemize}
\item the \textit{encoder} function $\mathbf{f}_{n}^E(\cdot; \boldsymbol{\theta}_E) : \mathbf{x} \mapsto \mathbf{\tilde{x}}_n = \mathbf{f}_{n}^E( \mathbf{x}; \boldsymbol{\theta}_E)$, where $\mathbf{f}_{n}^E (\cdot; \boldsymbol{\theta}_E) : \mathbb{R}^{N} \rightarrow \mathbb{R}^{n}$ and $n \ll N$, mapping the high-dimensional input $\mathbf{x}$ onto a low-dimensional code $\mathbf{\tilde{x}}_n$;
\item the \textit{decoder} function $\mathbf{f}^D(\cdot; \boldsymbol{\theta}_D) : \mathbf{\tilde{x}}_n \mapsto \mathbf{\tilde{x}} = \mathbf{f}^D(\mathbf{\tilde{x}}_n; \boldsymbol{\theta}_D)$, where $\mathbf{f}^D(\cdot; \boldsymbol{\theta}_D) : \mathbb{R}^{n} \rightarrow \mathbb{R}^{N}$, mapping the low-dimensional code $\tilde{\mathbf{x}}_n$ to an approximation of the original high-dimensional input $\mathbf{\tilde{x}}$. 
\end{itemize}} 
(AE) \cite{lecun1998gradient, hinton1994autoencoders}. More precisely, $\tilde{\mathcal{S}}$
takes the form
\begin{equation}
\label{eq:g}
\tilde{\mathcal{S}} =  \{ {\mathbf{f}}^D(\mathbf{u}_n(t; \boldsymbol{\mu}, {\boldsymbol{\theta}_{DF}}); \boldsymbol{\theta}_{D})
\; | \; \mathbf{u}_n(t; \boldsymbol{\mu}, \boldsymbol{\theta}_{DF}) \in {\mathbb{R}}^{n},    \ t \in [0, T)  \; \textnormal{and} \; \boldsymbol{\mu} \in \mathcal{P} \subset {\mathbb{R}}^{n_{\mu}} \} 
\end{equation}
where ${\mathbf{f}}^D(\cdot; {\boldsymbol{\theta}}_D) : {\mathbb{R}}^n \rightarrow {\mathbb{R}^{N}}$ consists in the decoder function of a convolutional AE. This latter results from the composition of several layers (some of which are convolutional), depending upon a vector ${\boldsymbol{\theta}}_{D}$ collecting all the corresponding weights and biases. 

As a matter of fact, the approximation $\mathbf{\tilde{u}}(t; \boldsymbol \mu) \approx \mathbf{u}(t; \boldsymbol \mu)$ provided by the DL-ROM technique is defined as 
\begin{equation}
\mathbf{\tilde{u}}(t; \boldsymbol \mu, \theta_{DF}, \theta_D) = {\mathbf{f}}^D({\boldsymbol{\phi}}_n^{DF}(t; \boldsymbol{\mu}, {{\boldsymbol{\theta}}_{DF}}); \boldsymbol{\theta}_{D}).
\label{eq:reconstructed_solution}
\end{equation}
The encoder function  of the convolutional AE can then be exploited to map the FOM solution associated to $(t, \boldsymbol{\mu})$ onto a low-dimensional representation 
\begin{equation}
\label{eq:encoder}
 {\mathbf{\tilde{u}}_n}(t; \boldsymbol{\mu}, \boldsymbol{\theta}_{E}) = {\mathbf{f}}_{n}^E(\mathbf{u}(t; \boldsymbol{\mu}); \boldsymbol{\theta}_{E});
\end{equation}
 $\mathbf{f}_{n}^E(\cdot; \boldsymbol{\theta}_E) : \mathbb{R}^{N} \rightarrow \mathbb{R}^n$ denotes  the encoder function, depending upon a vector ${\boldsymbol{\theta}}_E$ of parameters.

Computing the DL-ROM approximation of ${\bf u}(t; \muv_{test})$, for any possible $t \in (0,T)$ and $\muv_{test} \in \mathcal{P}$, corresponds to the testing  stage of a DFNN and of the decoder function of a convolutional AE; this does not require the evaluation of the encoder function. {We remark that our DL-ROM strategy overcomes the three major computational bottlenecks implied by the use of projection-based ROMs, since:
\begin{itemize}
	\item[-] the dimension of the DL-ROM can be kept extremely small;
			\item[-] the  time resolution required by the DL-ROM can be chosen to be  larger than the one required by the numerical solution of dynamical systems in cardiac electrophysiology; 
		\item[-] the DL-ROM can be queried at any desired time instant, without requiring the solution of a dynamical system until that time;
		\item[-] the DL-ROM does not require to account for the dynamics of the gating variables, thus avoiding any hyper-reduction stage. This advantage, already visible when employing a single gating variable as in the test cases addressed later in this paper, might become even more effective when dealing with more realistic ionic models (the so-called {\em I} and {\em II generation models}), when dozens of additional variables in the system of ODEs must be accounted for \cite{franzone2014mathematical}. 
\end{itemize}
} The training stage consists in solving the following optimization problem, in the  variable $\boldsymbol{\theta} = (\boldsymbol{\theta}_{E}, \boldsymbol{\theta}_{DF}, \boldsymbol{\theta}_{D})$, after the snapshot matrix ${\bf S}$ has been formed: 
\begin{equation}
\min_{\boldsymbol{\theta}} \mathcal{J}(\boldsymbol{\theta}) = \min_{\boldsymbol{\theta}} \frac{1}{N_s}\sum_{i=1}^{N_{train}} \sum_{k=1}^{N_t} \mathcal{L}(t^k, \boldsymbol \mu_i; \boldsymbol{\theta}), 
\label{eq:minimization_problem} 
\end{equation} 
where {$N_s = N_{train} N_t$} and  
\begin{equation}
\begin{split}
\mathcal{L}(t^k, \boldsymbol{\mu}_i;  {\boldsymbol{\theta}}) & = \frac{\omega_h}{2}\| \mathbf{u}(t^k; \boldsymbol{\mu}_i) - \mathbf{\tilde{u}}(t^k; \boldsymbol{\mu}_i,  {\boldsymbol{\theta}_{DF}, \boldsymbol{\theta}_D})\|^2 \\
& + \frac{1-\omega_h}{2}  \| \tilde{\mathbf{u}}_n(t^k; \boldsymbol{\mu}_i, \boldsymbol{\theta}_E) -  {\mathbf{u}}_n(t^k; \boldsymbol{\mu}_i,  {\boldsymbol{\theta}_{DF}})\|^2, 
\label{eq:loss_encoder}
\end{split}
\end{equation}
with $\omega_h \in [0,1]$. The \emph{per-example} loss function (\ref{eq:loss_encoder}) combines the reconstruction error (that is, the error between the FOM solution and the DL-ROM approximation) and the error between the  {intrinsic coordinates} and the output of the encoder. 

\begin{figure}[h!]
\centering
\includegraphics[scale=0.375]{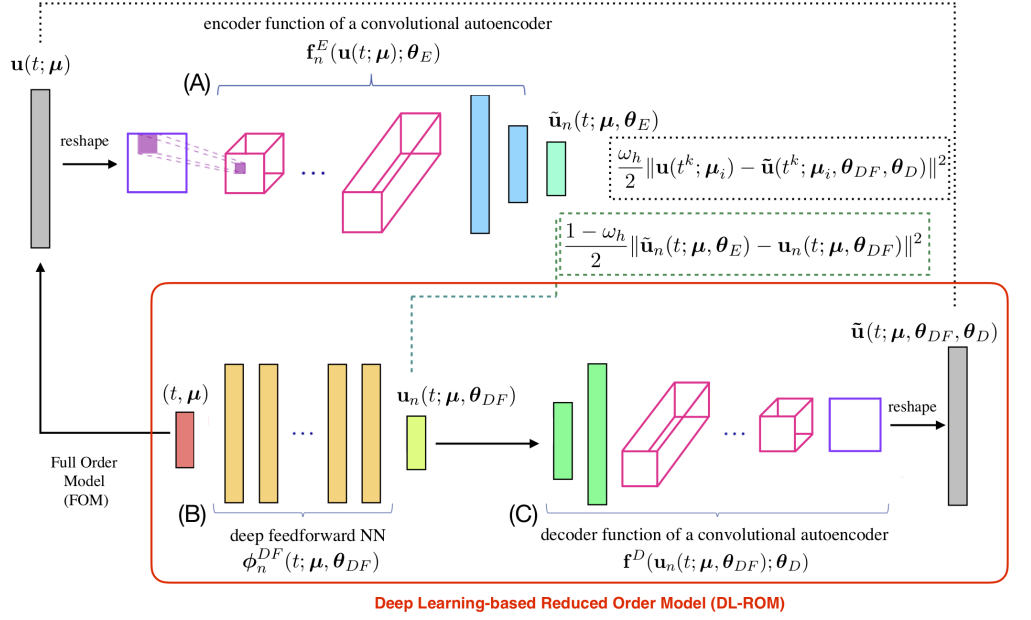}
\caption{{\bf DL-ROM architecture.} DL-ROM architecture used during the training phase. In the red box, the DL-ROM to be queried for any new selected couple $(t,\boldsymbol{\mu})$ during the testing phase. The FOM solution $\mathbf{u}(t; \boldsymbol{\mu})$ is provided as input to block (A) which outputs $\tilde{\mathbf{u}}_n(t; \boldsymbol{\mu})$. The same parameter instance associated to the FOM, i.e. $(t; \boldsymbol{\mu})$, enters block (B) which provides as output $\mathbf{u}_n(t; \boldsymbol{\mu})$ and the error between the low-dimensional vectors (dashed green box) is accumulated. The intrinsic coordinates $\mathbf{u}_n(t; \boldsymbol{\mu})$ are given as input to block (C) returning the ROM approximation $\mathbf{\tilde{u}}(t; \boldsymbol \mu)$. Then the reconstruction error (dashed black box) is computed.}
\label{fig:Fig1}
\end{figure}

The architecture of DL-ROM is the one shown in \figurename~\ref{fig:Fig1}. The encoder function is used only during the training and validation steps; it is instead discarded during the testing phase. See  \cite{fresca2020comprehensive} for further algorithmic details about the training and the testing algorithms required to build and evaluate a DL-ROM.

We highlight that the DL-ROM technique does not require to solve a (reduced) nonlinear dynamical system for the reduced degrees of freedom as in \eqref{eq:ROM-RB}; rather, it evaluates a nonlinear map for any given couple $(t, \muv_{test})$, for each $t \in (0,T)$. Numerical results are extremely accurate, the mean relative error is indeed below $1 \%$ (see, e.g., Test 2), even if the causality intrinsic to  the parabolic  nature of the diffusion-reaction equation providing the monodomain model is broken when computing the DL-ROM approximation. Moreover, the map features an extremely low dimension, in the most favorable scenario equal to $n_{\muv} + 1$. From a computational perspective, remarkable gains and simplifications can be obtained against a linear ROM, since {\em (i)} no hyper-reduction is required to enhance the evaluation of any $\muv$- or ${\bf u}$-dependent quantity, and {\em (ii)} even more interestingly, there is no need to evaluate the dynamics of the recovery variable $w$ if one is only interested in the electrical potential.  

\section*{Results and discussion}

We now assess the computational performances of the proposed DL-ROM strategy on three relevant test cases in cardiac electrophysiology. 
Our choice of the numerical tests is aimed at highlighting the performance of our DL-ROM method in challenging electrophysiology problems, namely pathological cases in portion of cardiac tissues or physiological scenarios on realistic left ventricle geometries.

 {The architecture used to perform all the numerical tests is the one reported in the \nameref{SI_Appendix}.} To solve the optimization problem (\ref{eq:minimization_problem})-(\ref{eq:loss_encoder}) we use the ADAM algorithm \cite{kingma2015adam} with a starting learning rate equal to $\eta = 10^{-4}$. Moreover, we perform cross-validation by splitting the data in training and validation and following a proportion 8:2 and we implement an early-stopping  regularization technique  to reduce overfitting \cite{goodfellow2016deep}.

To evaluate the performance of the DL-ROM, we use the loss function (\ref{eq:loss_encoder}) and on an   error indicator defined as 
\begin{equation}
\epsilon_{rel} = \frac{1}{N_{test}} \sum_{i  = 1}^{N_{test}} \left(\displaystyle \frac{\sqrt{ \sum_{k=1}^{N_t} || \mathbf{u}^k(\boldsymbol{\mu}_{test,i}) - \mathbf{\tilde{u}}^k(\boldsymbol{\mu}_{test,i}) ||^2}}{\sqrt{\sum_{k=1}^{N_t} || \mathbf{u}^k(\boldsymbol{\mu}_{test,i}) ||^2}} \right).
\label{eq:error_indicator}
\end{equation}

Neural networks required by our DL-ROM technique  have been implemented by means of the 	\texttt{Tensorflow} deep learning framework \cite{abadi2016tensorflow};  numerical simulations have been carried out on a workstation equipped with an Nvidia GeForce GTX 1070 8 GB GPU.

\subsection*{Test 1: Two-dimensional slab with ischemic region}
 
We consider the computation of the transmembrane potential in a square slab $\Omega = (0, 10 \; \textnormal{cm})^2$ of cardiac tissue in presence of an ischemic (non-conductive) region. The ischemic region may act as anatomical driver of cardiac arrhythmias like tachycardias and fibrillations. The system we want to solve is a slight modification of equations (\ref{eq:monodomain}), accounting for the presence of a non-conductive region which  affects both the conductivity tensor and the ionic current term.  
The ischemic portion of the domain is modeled by replacing the conductivity tensor $\bf{D}(\mathbf{x})$, defined in (\ref{eq:conductivity_tensor}), with $\bf{\bar{D}}(\mathbf{x}; \boldsymbol \mu) = \sigma(\mathbf{x}, \boldsymbol \mu) \bf{D}(\mathbf{x})$, where the function $\sigma(\mathbf{x}, \boldsymbol \mu)$ is given by 
\begin{equation}
\label{eq:rho}
\begin{aligned}
& \sigma(\mathbf{x}; \boldsymbol \mu) = \rho(\mathbf{x}; \boldsymbol \mu) + \sigma_0 (1 - \rho(\mathbf{x}; \boldsymbol \mu)), \medskip \\ 
& \rho(\mathbf{x}; \boldsymbol \mu)= 1 - \exp \bigg(- \frac{(x_1 - \mu_1)^4 + (x_2 - \mu_2)^4}{2\alpha^2} \bigg).
\end{aligned}
\end{equation}
In this case,  $n_{\mu}=2$ parameters are considered, representing the coordinates of the center of the scar, belong to the parameter space $\mathcal{P}= {[3.5, 6.5 \; \textnormal{cm}]}^2$. 
Moreover, $\alpha = 7$ cm\textsuperscript{2}, $\sigma_0 = 10^{-4}$, the transversal and longitudinal conductivities are  $\sigma_t = 12.9 \cdot 0.1$ cm\textsuperscript{2}/ms and $\sigma_l = 12.9 \cdot 0.2$ cm\textsuperscript{2}/ms, respectively, and $\mathbf{f}_0 = (1, 0)^T$, meaning that the tissue fibers are parallel to the $x-$axis. Similarly, the ionic current $I_{ion}(u, w)$ in (\ref{eq:monodomain}) is replaced by 
$\bar{I}_{ion}(u, w; \boldsymbol{\mu})= \rho(\mathbf{x}; \boldsymbol{\mu}) I_{ion}(u, w)$. 
The applied current takes the form
\begin{equation*}
I_{app}(\mathbf{x}, t) = C \exp \bigg( -\frac{||\mathbf{x}||^2}{\beta} \bigg) \mathbf{1}_{[0, \bar{t}]}(\tilde{t}),
\end{equation*}
where $C = 100$ mA, $\beta = 0.02$ cm\textsuperscript{2} and $\bar{t} = 2$ ms. The parameters appearing in (\ref{eq:aliev_panfilov}) are set to $K = 8$, $a = 0.01$, $b = 0.15$, $\varepsilon_0 = 0.002$, $c_1 = 0.2$, and $c_2 = 0.3$, see \cite{goktepe2010atrial}. The equations have been discretized in space through linear finite elements by considering $N = 64 \times 64 = 4096$ grid points. For the time discretization and the treatment of nonlinear terms, we use a one-step, semi-implicit, first order scheme (see \cite{pagani2018numerical} for further details) by considering a time step  $\Delta t = 0.1/12.9$ over  $(0,T)$ with $T=400$ ms. 

{For the training phase, we uniformly sample $N_t = 1000$ time instances over $(0,T)$ and consider $N_{train} = 49$ training-parameter instances, with $\boldsymbol{\mu}_{train}=(3.5 + i 0.5, 3.5 + j 0.5)$, $i,j = 0, \ldots, 6$.} The maximum number of epochs is set equal to $N_{epochs} = 10000$, the batch size is $N_b = 40$ and, regarding the early-stopping criterion, we stop the training if the loss function does not decrease in 500 epochs.  {For the testing phase,  $N_{test} = 36$ testing-parameter instances $\boldsymbol{\mu}_{test}=(3.75 + i 0.5, 3.75 + j 0.5)$, $i,j = 0, \ldots, 5$, have been considered.}

\begin{figure}[h!t]
\centering
\includegraphics[scale=0.475]{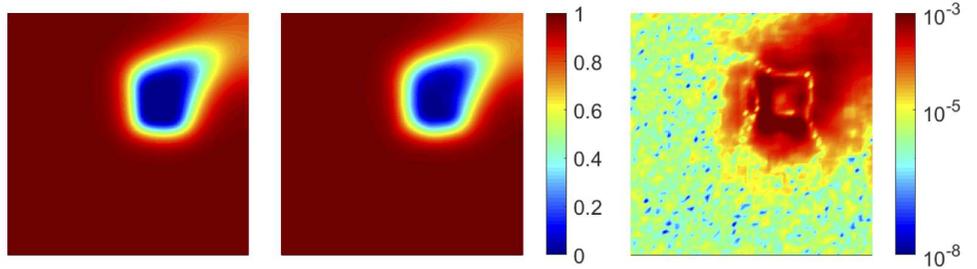}
\caption{{\bf Test 1: comparison between FOM and DL-ROM solutions for a testing-parameter instance.}
FOM solution (left), DL-ROM solution with $n=3$ (center)  and relative error $\boldsymbol{\epsilon}_k$ (right) for the testing-parameter instance $\boldsymbol{\mu}_{test} = (6.25, 6.25)$ cm at $\tilde{t} = 100$ ms. The maximum of the relative error $\boldsymbol{\epsilon}_k$ is $10^{-3}$ and it is associated to the diseased tissue.}
\label{fig:Fig2}
\end{figure}

\begin{figure}[h!t]
\centering
\includegraphics[scale=0.475]{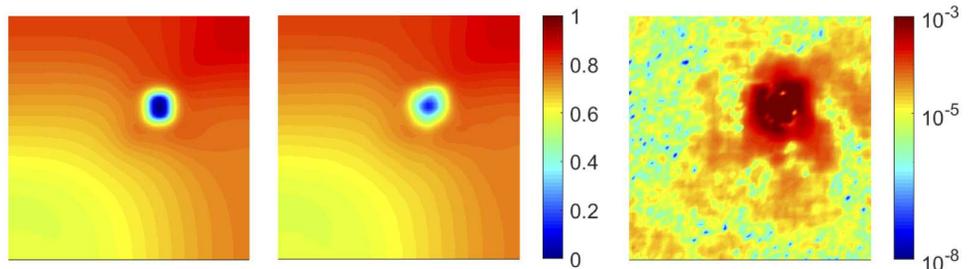}
\caption{{\bf Test 1: comparison between FOM and DL-ROM solutions for a testing-parameter instance.}
FOM solution (left), DL-ROM solution with $n=3$ (center) and relative error $\boldsymbol{\epsilon}_k$ (right) for the testing-parameter instance $\boldsymbol{\mu}_{test} = (6.25, 6.25)$ cm at $\tilde{t} = 356$ ms. The maximum of the relative error $\boldsymbol{\epsilon}_k$ is $10^{-3}$ and it is associated to the diseased tissue.}
\label{fig:Fig3}
\end{figure}

In \figurenames~\ref{fig:Fig2} and \ref{fig:Fig3} we show the FOM  and the DL-ROM solutions, the latter obtained with $n = 3$ for the testing-parameter instance $\boldsymbol{\mu}_{test}=(6.25, 6.25)$ cm at $\tilde{t} = 100$ and 356 ms, respectively, together with the relative error $\boldsymbol{\epsilon}_k \in \mathbb{R}^{N}$, for $k = 1, \ldots, N_t$, defined as 
\begin{equation}
\displaystyle \boldsymbol{\epsilon}_k = \displaystyle \frac{ | \mathbf{u}^k(\boldsymbol{\mu}_{test}) - \mathbf{\tilde{u}}^k(\boldsymbol{\mu}_{test}) |}{\sqrt{\frac{1}{N_t}\sum_{k=1}^{N_t} || \mathbf{u}^k(\boldsymbol{\mu}_{test}) ||^2}}.
\label{eq:relative_error}
\end{equation}
While (\ref{eq:error_indicator}) is a synthetic indicator, the quantity defined in \eqref{eq:relative_error}   is instead a function of the space independent variable. In \figurename~\ref{fig:Fig2} the tissue is depolarized except for the region occupied by scar and surrounding it, which is clearly characterized by a slower conduction. In \figurename~\ref{fig:Fig3} the tissue is starting to repolarize and even if the shape of the ischemic region is not sharply reproduced, the DL-ROM solution is able to capture the diseased (non-conductive) nature of this portion of tissue.

\begin{figure}[h!t]
\centering
\includegraphics[scale=0.2]{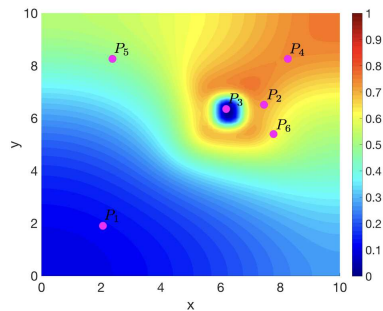}
\smallskip
\caption{{\bf Test 1: location of points $\mathbf{P_i}$.}
FOM solution evaluated for $\boldsymbol{\mu}_{test} = (6.25, 6.25)$ cm at $\tilde{t} = 400$ ms together with the points $P_1,\ldots, P_6$.}
\label{fig:Fig4}
\end{figure}

\begin{figure}[!ht]
\centering
\includegraphics[scale=0.115]{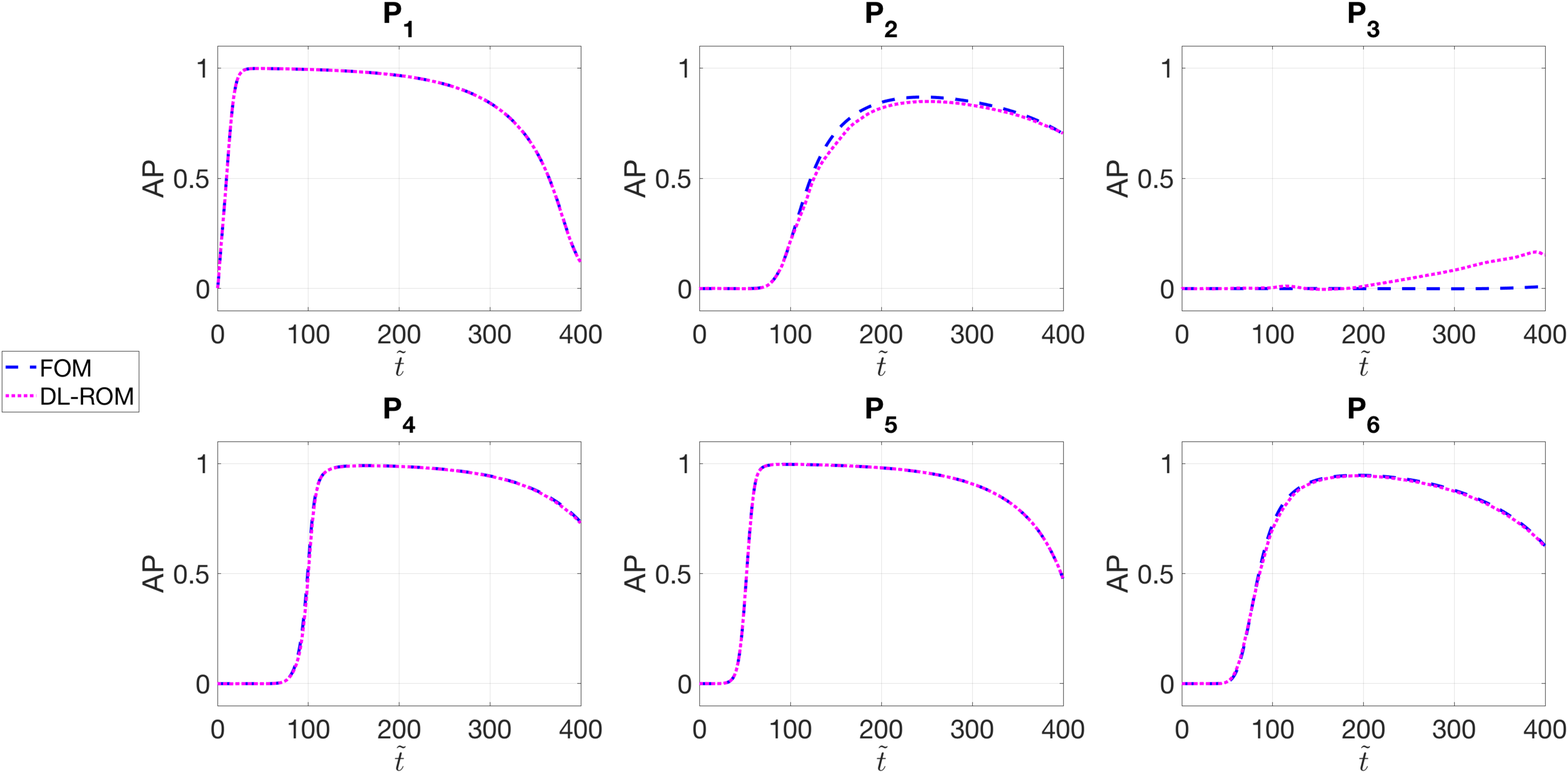}
\smallskip
\caption{{\bf Test 1: comparison between the FOM and DL-ROM APs at $\mathbf{P_i}$.}
APs evaluated  for 
$\boldsymbol{\mu}_{test} = (6.25, 6.25)$ cm at points $P_1,\ldots, P_6$. The DL-ROM, with $n=3$, is able to to sharply reconstruct the AP in almost all the points and the main features are captured also for the points close to the scar.}
\label{fig:Fig5}
\end{figure}

In \figurename~\ref{fig:Fig5} we show the action potentials (APs) computed at the six points $P_1, \ldots, P_6$ reported in \figurename~\ref{fig:Fig4}. The DL-ROM is able to provide an accurate reconstruction of the AP at almost all points; the maximum error is associated to the point $P_3$, the closest one to the center of the scar, for $\tilde{t} \ge 200$ ms. However,  even in this case, the DL-ROM technique is able to capture the difference, in terms of AP values, between the  diseased and the healthy tissue. 

\begin{figure}[!ht]
\centering
\includegraphics[scale=0.12]{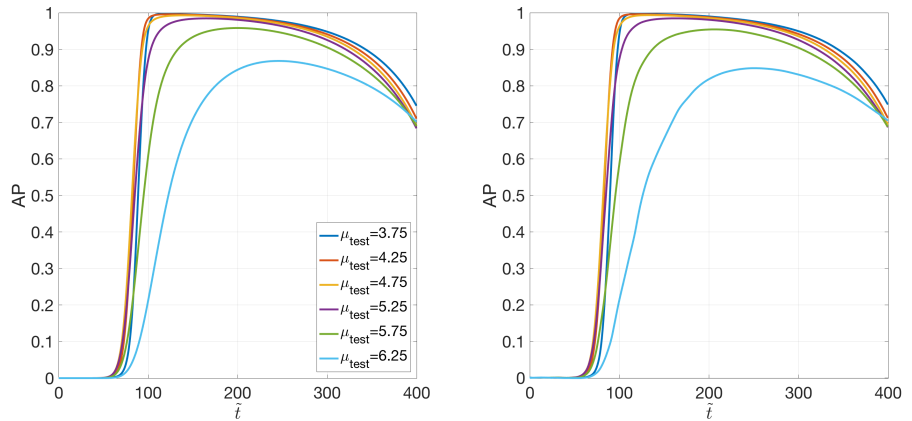}
\smallskip
\caption{{\bf Test 1: variability of the FOM and DL-ROM solutions over the parameter space.}
FOM (right) and DL-ROM (left) AP variability over $\mathcal{P}$ at $P = (7.46, 6.51)$ cm. The DL-ROM sharply reconstructs the FOM variability over $\mathcal{P}$.}
\label{fig:Fig6}
\end{figure}

The AP variability  across the parameter space   characterizing both the FOM and the DL-ROM solutions is shown in  \figurename~\ref{fig:Fig6}. Still with a DL-ROM  dimension $n = 3$, we report the APs for $\boldsymbol{\mu}_{test} = (\mu_{test}, \mu_{test})$ cm, with $\mu_{test} = 3.75, 4.25, 4.75, 5.25, 5.75, 6.25$, evaluated at  $P = (7.46, 6.51)$ cm. The DL-ROM  is able to capture such variability over $\mathcal{P}$; moreover, the larger $\boldsymbol{\mu}_{test}$, the  smaller the distance between the point $P$ and the scar, with their proximity impacting on the shape and the values of the AP.  {In particular, for $\mu_{test} = 6.25$,  the point $P$ falls into the {\em grey zone}.}

By using the DL-ROM technique and setting the dimension of the nonlinear trial manifold equal to the dimension of the solution manifold, i.e. $n = 3$, we obtain an error indicator (\ref{eq:error_indicator}) of $\epsilon_{rel} = 2.01 \cdot 10^{-2}$. In order to assess the computational efficiency of DL-ROM, we  compare it with the POD-Galerkin  ROM relying on $N_c$ local reduced bases; we report in \tablename~\ref{tb:Tb1} the maximum and minimum number of basis functions, among all the clusters, required by the POD-Galerkin ROM \cite{quarteroni2016reduced, pagani2018numerical} to achieve the same accuracy. 

\begin{table}[!ht]
\centering
\caption{
{\bf Test 1: dimensions of the POD-Galerkin ROM linear trial manifolds by varying the number of clusters.}}
\begin{tabular}{|l|l|l|l|}
\hline
$N_c = 1$ & $N_c = 2$ & $N_c =4$ & $N_c = 6$ \\ \hline
250 & 219 & 200 & 193 \\ \hline
 & 107 & 35 & 26 \\ \hline
\end{tabular}
\begin{flushleft} Maximum and minimum dimensions of the local  reduced bases (that is, linear trial manifolds) built by the  POD-Galerkin  ROM for different numbers $N_c$ of clusters.
\end{flushleft}
\label{tb:Tb1}
\end{table}

\begin{figure}[h!t]
\centering
\includegraphics[scale=0.23]{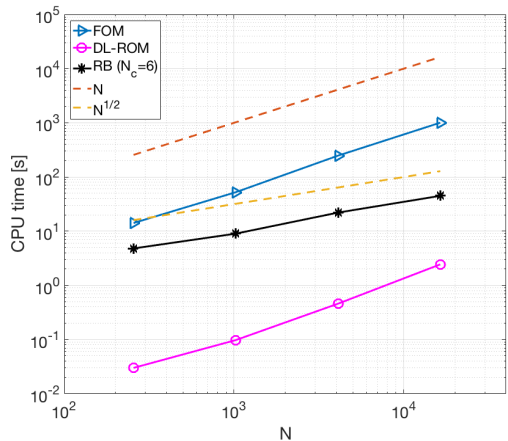}
\vspace{0.1cm}
\caption{{\bf Test 1: FOM, DL-ROM and POD-Galerkin ROM CPU computational times.}
CPU time required to solve the FOM, by DL-ROM at testing time with $n = 3$ and by the  POD-Galerkin  ROM at testing time with $N_c = 6$ vs. $N$. The DL-ROM provides the smallest testing computational time for each $N$ considered.}
\label{fig:Fig7}
\end{figure}

In \figurename~\ref{fig:Fig7} we compare the CPU time required to solve the FOM (through linear finite elements) over the time interval $(0, T)$, with the one needed by DL-ROM with $n = 3$, and the POD-Galerkin ROM  with $N_c = 6$ local reduced bases, at testing time, by varying the FOM dimension $N$. Here, with testing time we refer, both for the DL-ROM and the POD-Galerkin ROM, to the time needed to query 
the ROM over the whole interval $(0, T)$, by using for each technique the proper time resolution, 
 for a given testing-parameter instance. Since the DL-ROM solution can be queried at a given time without requiring any solution of a dynamical system to recover the former time instances, the DL-ROM can employ larger time windows compared to the time steps  required by the solution of the FOM and  POD-Galerkin ROM dynamical systems for the cases at hand. This fact also has a positive impact on the data used during the training phase\footnote{Indeed, in order to build the snapshot matrix, we uniformly sample $N_t$ time instances of the FOM solution over $T/\Delta t = 4000$ time steps; for each training parameter instance, only $25\%$ of 4000 snapshots are retained from the FOM solution in the DL-ROM case, against  $4000$ snapshots in the POD-Galerkin ROM case.}. The speed-up obtained, for each value of $N$ considered, is reported in \tablename~\ref{tb:Tb1}. Both the DL-ROM and the POD-Galerkin ROM allow us to decrease the computational costs associated to the computation of the FOM solution for a  testing-parameter instance. However, for a desired level of accuracy, CPU times required by the POD-Galerkin ROM during the testing phase are remarkably higher than  the ones required by a DL-ROM with $n = 3$.

Both the DL-ROM and the POD-Galerkin  ROM depend on the FOM dimension $N$. In the case of DL-ROM, the dependency on $N$ at testing time, for a fixed value of $\Delta t$, is due to the presence of the decoder function; indeed, the process of learning the reduced dynamics (and so  the dimension of the nonlinear trial manifold) does not depend on the FOM dimension. On the other hand, the dependence of the POD-Galerkin  ROM on the FOM dimension  also impacts on the dimension of the local linear trial manifolds: in general, by increasing $N$ the dimension of each local linear subspace also increases. Referring to \figurename~\ref{fig:Fig7} and \tablename~\ref{tb:Tb2}, the CPU time required by the DL-ROM {at testing time} scales linearly with  $N$, instead the one required by the POD-Galerkin ROM scales linearly with  $\sqrt{N}$. In particular, even for the larger FOM dimension considered ($N = 16384$ for this test case), our DL-ROM is 19 times faster than the POD-Galerkin ROM. We are not able to run simulations for $N > 16384$, because of the limitation of the computing resources we have at our disposal. Despite the trend in \figurename~\ref{fig:Fig7} is apparently not favorable for the DL-ROM technique, practice indicates that the CPU time for DL-ROM is smaller than the one for the POD-Galerkin ROM  for {\em small} values of $N$, in other words only with very large values of $N$  the POD-Galerkin ROM   {outperforms} the DL-ROM strategy. Indeed, a linear fitting of the DL-ROM and the POD-Galerkin ROM CPU times\footnote{$N = 65536$ and $N = 262144$ for this test case represent FOM dimensions corresponding to  mesh sizes $h$ needed 
to solve, by means of linear finite elements,  the problem on a 3D slab geometry 
 both for physiological and pathological electrophysiology in the case a ten Tusscher-Panfilov ionic model \cite{ten2006alternans} is used. This latter would indeed require smaller values of $h$ compared to the Aliev-Panfilov model, due to the shape of the AP. See, e.g.,  \cite{trayanova2011whole, plank2008from} for further details.}
 in \figurename~\ref{fig:Fig7} highlights that  for $N = 65536$ and $N = 262144$, DL-ROM could be almost 10 and 5 times, respectively, faster than  the POD-Galerkin ROM for the same values of $N$. 
Note that the results of this section have been obtained by employing  the DL-ROM  on a single CPU, an architecture which is not favorable to neural networks\footnote{Indeed, all tests are performed on a node (20 Intel\textsuperscript{\textregistered} Xeon\textsuperscript{\textregistered} E5-2640 v4 2.4GHz cores), using 5 cores, of our in-house HPC cluster.}.  Further improvements are expected when employing our DL-ROM on a GPU for a given testing-parameter instance. 
\begin{table}[!ht]
\centering
\caption{
{\bf Test 1: DL-ROM and POD-Galerkin ROM vs. FOM speed-up.}}
\begin{tabular}{|l|l|l|l|l|}
\hline
 & $N = 256$ & $N = 1024$ & $N = 4096$ & $N = 16384$ \\ \hline
FOM vs. DL-ROM & 472 & 536 & 539 & 412 \\ \hline 
FOM vs. POD-Galerkin ROM & 3 & 6 & 12 &  22 \\ \hline  
\end{tabular}
\begin{flushleft} DL-ROM and POD-Galerkin ROM vs. FOM speed-up by varying $N$. The DL-ROM speed-up is remarkably higher than the one obtained by using the POD-Galerkin ROM.
\end{flushleft}
\label{tb:Tb2}
\end{table}

\begin{figure}[!ht]
\centering
\includegraphics[scale=0.14]{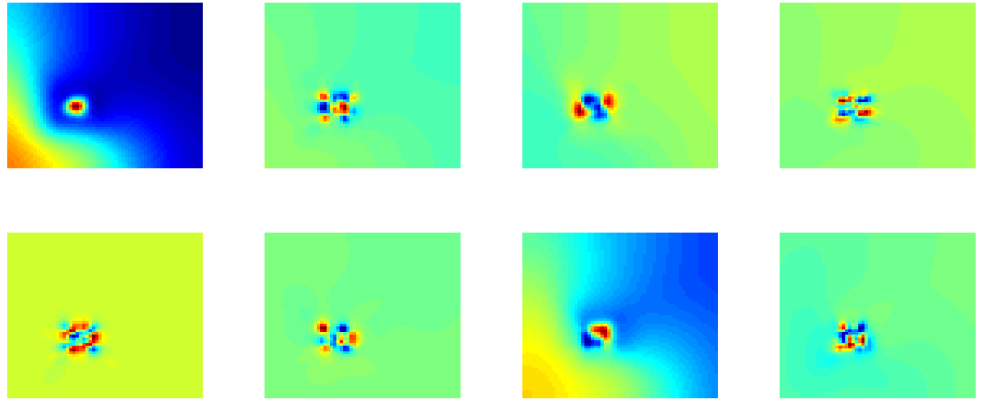}
\caption{{\bf Test 1: activations of the first convolutional layer of the encoder function for a testing-parameter instance.}
Feature maps of the first convolutional layer of the encoder function in the DL-ROM neural network for the testing-parameter instance $\boldsymbol{\mu}_{test} = (3.75, 3.75)$ cm at $\tilde{t} = 0.2$ ms.}
\label{fig:Fig8}
\end{figure}

\begin{figure}[!ht]
\centering
\includegraphics[scale=0.14]{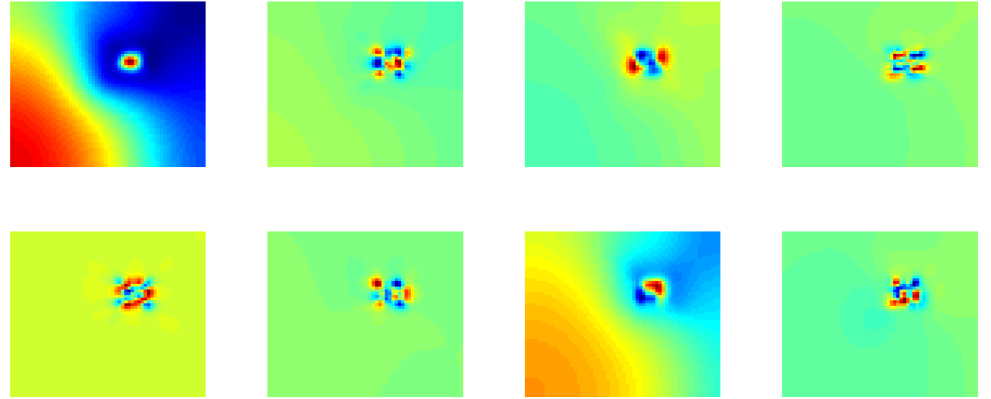}
\caption{{\bf Test 1: activations of the first convolutional layer of the encoder function for a testing-parameter instance.}
Activations of the first convolutional layer of the encoder function in the DL-ROM neural network for the testing-parameter instance $\boldsymbol{\mu}_{test} = (6.25, 6.25)$ cm at $\tilde{t} = 0.2$ ms.}
\label{fig:Fig9}
\end{figure}

In \figurenames~\ref{fig:Fig8} and \ref{fig:Fig9} we show the feature maps of the first convolutional layer of the encoder function $\sigma_1(W_1^k * \mathbf{u}^1(\boldsymbol{\mu_{test}}) + b_1^k)$, for $k = 1, \ldots, 8$, in the DL-ROM neural network when the FOM solution for the testing-parameter instances $\boldsymbol{\mu}_{test} = (3.75, 3.75)$ cm and $\boldsymbol{\mu}_{test} = (6.25, 6.25)$ cm at $t = 0.2$ ms, are  provided as inputs. At this stage, the feature maps retain most of the information present in the FOM solution. Moreover, by considering the two testing-parameter instances, we observe the translation equi-variance property \cite{goodfellow2016deep} that convolutional layers hold when applied to the part of cardiac tissue corresponding to the scar.  {Moving to deeper layers, feature maps become increasingly abstract, and less visually interpretable; however, the extracted high-level features are still related both to the ischemic region and the electrical activation pattern.}

\begin{figure}[ht]
\centering
\includegraphics[scale=0.22]{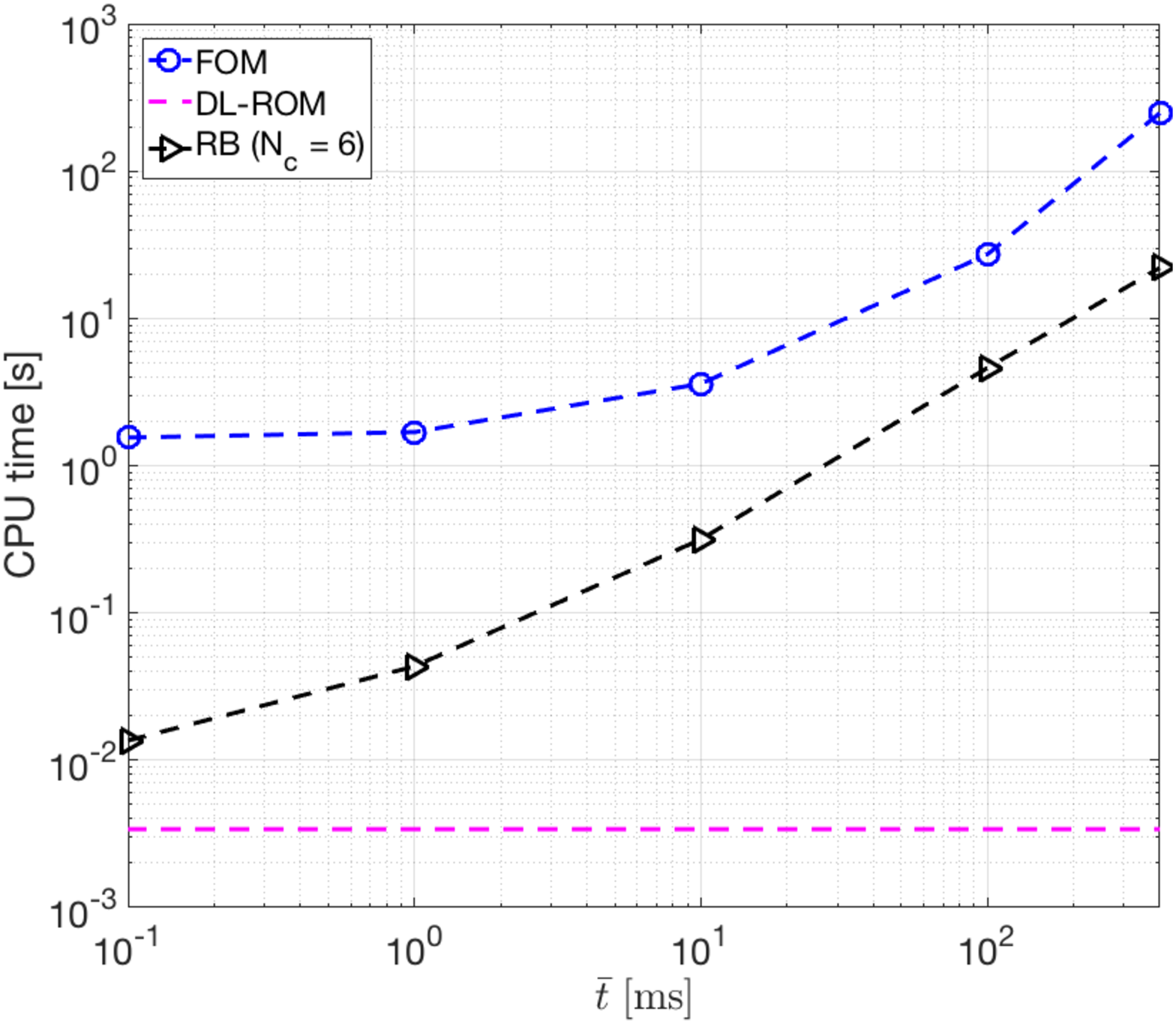}
\smallskip
\caption{{\bf Test 1: FOM, POD-Galerkin ROM and DL-ROM CPU computational times.} FOM, POD-Galerkin ROM and DL-ROM CPU computational times to compute $\tilde{\mathbf{u}}(\bar{t}; \boldsymbol{\mu}_{test})$ vs. $\bar{t}$ averaged over the testing set. Thanks to the fact that the DL-ROM can be queried at any time istance it is extremely efficient in computing $\tilde{\mathbf{u}}(\bar{t}; \boldsymbol{\mu}_{test})$ with respect to both the FOM and the POD-Galerkin ROM.}
\label{fig:Fig10}
\end{figure}

{
We highlight that since the DL-ROM solution can be evaluated at any desired time instance without {\em solving} any dynamical system, the resulting computational time entailed by the DL-ROM at testing time are drastically reduced compared to the ones required by the  FOM or the POD-Galerkin ROM to compute solutions at a particular time instance. In \figurename~\ref{fig:Fig10} we show the DL-ROM, FOM and POD-Galerkin ROM CPU time needed to compute the approximated solution at $\bar{t}$, for $\bar{t} =$ 1, 10, 100 and 400 ms averaged over the testing set and with $N = 4096$. We perform the training phase of the POD-Galerkin ROM over the original time interval $(0, T)$ ms and we report the results for $N_c =6$, the number of clusters for which the smallest computational time is obtained. The DL-ROM CPU time to compute $\tilde{\mathbf{u}}(\bar{t}; \boldsymbol{\mu}_{test})$ does not vary over $\bar{t}$ and, by choosing $\bar{t} = T$, the DL-ROM speed-ups are equal to $7.3 \times 10^{4}$ and $6.5 \times 10^{3}$ with respect to the FOM and the POD-Galerkin ROM, with $N_c=6$, computational times.}

\subsection*{Test 2: Two-dimensional slab with figure of eight re-entry}

The most recognized cellular mechanisms sustaining atrial tachycardia is re-entry \cite{nattel2002new}. The particular kind of re-entry we deal with in this test case is called {\em figure of eight} re-entry, and can be obtained by solving equations (\ref{eq:monodomain}). To induce the re-entry, we apply a classical S1-S2 protocol \cite{kunisch2012optimal,franzone2014mathematical}. In particular, we consider a square slab of cardiac tissue $\Omega = (0, 2 \; \textnormal{cm})^2$ and apply an initial stimulus at the bottom edge of the domain, i.e. 
\begin{equation}
I_{app}^1(\mathbf{x}, t) = \mathbf{1}_{\Omega_1}(\mathbf{x})\mathbf{1}_{[{t}_1^{i}, {t}_1^{f}]}(\tilde{t}),
\label{eq:I_app_1}
\vspace{-0.1cm}
\end{equation}
where $\Omega_1 = \{ \mathbf{x} \in \Omega : y 	\le 0.1 \}$, $t
_1^{i} = 0$ ms and $t_1^{f} = 5$ ms. 

\noindent A second stimulus under the form 
\begin{equation}
I_{app}^2(\mathbf{x}, t; \mu) = \mathbf{1}_{\Omega_2(\mu)}(\mathbf{x})\mathbf{1}_{[t_2^{i}, t_2^{f}]}(\tilde{t}),
\label{eq:I_app_2}
\end{equation}
with $\Omega_2(\mu) = \{ \mathbf{x} \in \Omega : (x - 1)^2 + (y - \mu)^2 \le (0.2)^2\}$, $t_2^{i} = 70$ ms and $t_2^{f} = 75$ ms, is then applied. The parameter $\mu$, consisting in the $y$-coordinate of the center of the second circular stimulus, ranges in the parameter space $\mathcal{P} = [0.8, 1.1]$ cm. This choice has been made to obtain a re-entry elicited and sustained until $T = 175$ ms. 

We restrict our study to the time interval [95, 175] ms, i.e. we do not consider the first time instances in which the re-entry has not   arisen yet, being them equal over $\mathcal{P}$. The time step  is $\Delta t = 0.2/12.9$. We consider $N = 256 \times 256 = 65536$ grid points, implying a mesh size $h = 0.0784$ mm; this mesh size is recognized to correclty solve the tiny transition front developing during depolarization of the tissue, as highlighted in \cite{trayanova2011whole, plank2008from}. The fibers are parallel to the $x$-axis and the conductivities in the longitudinal and transversal directions to the fibers are $\sigma_l = 2 \times 10^{-3}$ cm\textsuperscript{2}/ms and $\sigma_t = 3.1 \times 10^{-4}$ cm\textsuperscript{2}/ms, respectively. The parameters appearing in (\ref{eq:aliev_panfilov}) are set to $K = 8$, $a = 0.1$, $b = 0.1$, $\varepsilon_0 = 0.01$, $c_1 = 0.14$, and $c_2 = 0.3$, see \cite{ten2004spiral}. 

The snapshot matrix is built by solving problem (\ref{eq:monodomain}), completed with the applied currents \ref{eq:I_app_1} and \ref{eq:I_app_2}, by means of linear finite elements and a semi-implicit scheme, over $N_t = 400$ time instances. Moreover, we consider $N_{train} = 13$ training-parameter instances uniformly distributed in the parameter space and $N_{test} = 12$ testing-parameter instances, each of them corresponding to the midpoint of two consecutive training-parameter instances. The maximum number of epochs is set equal to $N_{epochs} = 6000$, the batch size is $N_b = 3$, due to the high GPU memory occupation of each sample. Regarding the early-stopping criterion, we stop the training if the loss does not decrease in 1000 epochs. 

\begin{figure}[!ht]
\centering
\includegraphics[scale=0.425]{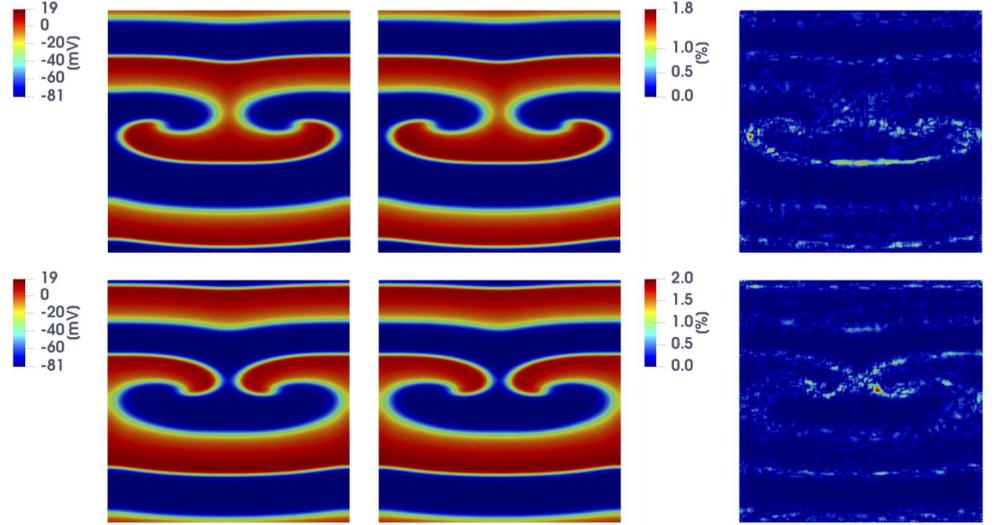}
\caption{{\bf Test 2: comparison between FOM and DL-ROM solutions for a testing-parameter instance.} FOM solution (left), DL-ROM one (center) with $n = 5$, and relative error $\boldsymbol{\epsilon}_k^s$ (right) at $\tilde{t} = 141.2$ ms (top) and $\tilde{t} = 157.2$ ms (bottom), for the testing-parameter instance $\mu_{test} = 0.9625$ cm. The relative error $\boldsymbol{\epsilon}_k^s$ is below the 2$\%$ both for $\tilde{t} = 141.2$ ms and $\tilde{t} = 157.2$ ms, the maximum value of the error being associated to very few points of the domain.}
\label{fig:Fig11}
\end{figure}

In \figurename~\ref{fig:Fig11} we show the FOM solution and the DL-ROM one obtained by setting the reduced dimension to $n = 5$, for the testing-parameter instance $\mu_{test} = 0.9625$ cm, at $\tilde{t} = 141.2$ ms and $\tilde{t} = 157.2$ ms, together with the relative error $\boldsymbol{\epsilon}_k^s \in \mathbb{R}^{N}$, for $k = 1, \ldots, N_t$, defined as  
\begin{equation}
	{\boldsymbol{\epsilon}_k^s = \displaystyle \frac{ | \mathbf{u}^k({\mu}_{test}) - \mathbf{\tilde{u}}^k({\mu}_{test}) |}{
	\|\mathbf{u}^k({\mu}_{test})\|_1
	}.}
\label{eq:relative_error_space}
\end{equation}

\begin{figure}[!ht]
\centering
\includegraphics[scale=0.125]{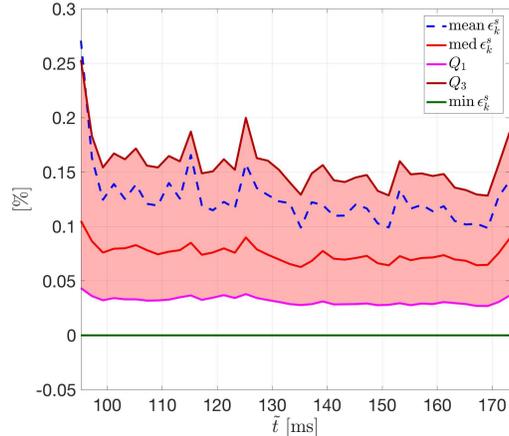}
\caption{{\bf Test 2: trend of the relative error over time.} Relative error $\boldsymbol{\epsilon}_{k}^s$ vs. $\tilde{t}$ with $n = 5$ for the testing-parameter instance $\mu_{test} = 0.9625$ cm (the red band indicates the IQR). The distribution of the error maintains uniform over time. 
}
\label{fig:Fig12}
\end{figure}

The trend of the relative error (\ref{eq:relative_error_space}) over time, for the selected testing-parameter instance $\mu_{test} = 0.9625$ cm, is depicted in \figurename~\ref{fig:Fig12}; we highlight that the error is always smaller than 1\%. In particular, in \figurename~\ref{fig:Fig12} we show the mean (over the domain), the median, and the first and third quartile of the relative error, as well as its minimum. The interquartile range (IQR) shows that the distribution of the error is almost uniform over time, and that  the maximum error is associated to the first time instant -- this latter being  the time instant at which the solution is most different over $\mathcal{P}$. 

\begin{figure}[!ht]
\centering
\includegraphics[scale=0.425]{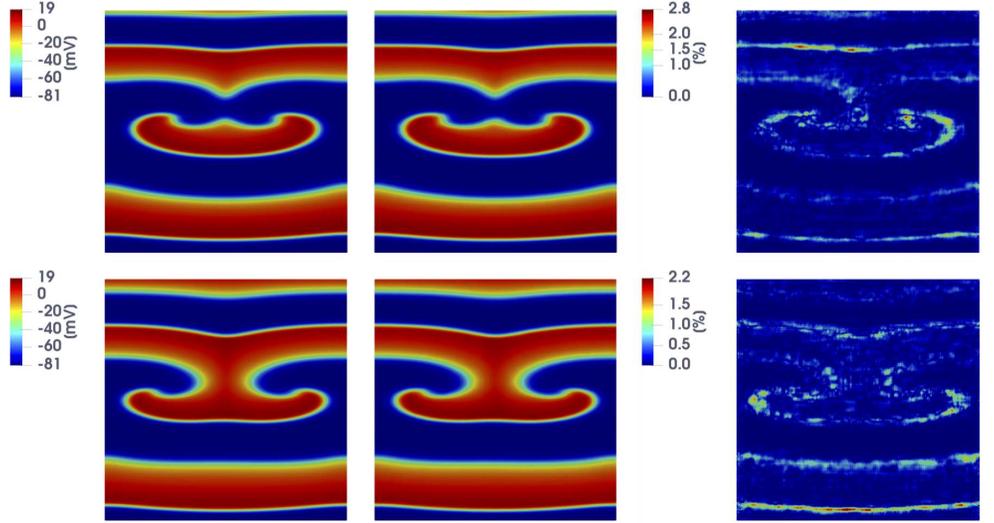}
\caption{{\bf Test 2: comparison between FOM and DL-ROM solutions for different testing-parameter instances.} FOM solution (left), DL-ROM one (center) with $n = 5$, and relative error $\boldsymbol{\epsilon}_k^s$ (right) at $\tilde{t} = 175$ ms, for the testing-parameter instance $\mu = 0.8125$ cm (top) and $\mu = 1.0625$ cm (bottom). The relative error $\boldsymbol{\epsilon}_k^s$ is below the 2.8$\%$ both for $\mu = 0.8125$ cm and $\mu = 1.0625$ cm, the maximum value of the error being associated to very few points of the domain.}
\label{fig:Fig13}
\end{figure}

In \figurename~\ref{fig:Fig13} we show the FOM and the DL-ROM solutions, the latter obtained by setting $n = 5$, for the last time instance, i.e. at $\tilde{t} = 175$ ms, for $\mu_{test} = 0.8125$ cm and $\mu_{test} = 1.0625$ cm, in order to point out the variability of the solution over the parameter space and the ability of DL-ROM to capture it.

We now compare the  computational times required by the FOM, the POD-Galerkin ROM (for different values of $N_c$) and the DL-ROM, keeping for all the same degree of accuracy achieved by DL-ROM, i.e. $\epsilon_{rel} = 7.87 \times 10 ^{-3}$, and running the code on the hardware each implementation is optimized for -- a CPU for the FOM and the POD-Galerkin ROM, a GPU\footnote{Indeed, at each layer of a neural network thousands of identical computations must be performed. The most suitable hardware architectures to carry out this kind of operations are  GPUs because (i) they have more computational units (cores) and (ii) they have a higher bandwidth to retrieve from memory. Moreover, in applications requiring image processing, as CNNs, the graphics specific capabilities can be further exploited to speed up calculations.} for the DL-ROM. In \tablename~\ref{tb:Tb3} we report the CPU time needed to compute the FOM solution, 
and the POD-Galerkin ROM (at the testing phase), both on a full 64 GB node (20 Intel\textsuperscript{\textregistered} Xeon\textsuperscript{\textregistered} E5-2640 v4 2.4GHz cores),
and the GPU time required by the DL-ROM   to compute 875 time instances  {(the same number of time instants considered in the solution of the dynamical systems associated to the FOM and the POD-Galerkin ROM)} at testing time, by fixing its dimension to $n = 5$, on an Nvidia GeForce GTX 1070 8 GB GPU. For the sake of completeness, we also report  the computational time required by the DL-ROM  when employing a single CPU node. 
It is evident that a POD-Galerkin ROM, built employing a global reduced basis $(N_c=1)$, is not amenable to a complex and challenging pathological cardiac electrophysiology problem like the figure of eight re-entry. Using a nonlinear approach, for which the solution manifold is approximated through a piecewise linear trial manifold (as in the case of $N_c = 2$ or $N_c = 4$ local reduced bases) reduces the online computational time. However, the DL-ROM still confirms to provide a more efficient ROM,   almost 5 (or 2) times faster on the CPU, and 39 (or  19) faster on the GPU,  than the  POD-Galerkin ROM with $N_c = 2$ (or $N_c = 4$) local reduced bases. 
\begin{table}[!ht]
\centering
\caption{\bf
{Test 2: FOM, POD-Galerkin ROM and DL-ROM computational times.}}
\begin{tabular}{|l|l|l|}
\hline
 & time [s] & FOM/ROM dimensions \\ \hline
FOM (CPU) &  382 & $N = 65536$ \\ \hline
DL-ROM (CPU/GPU) & 15/1.2 & $n = 5$ \\ \hline
POD-Galerkin ROM $N_c=1$ (CPU) & 103 & $n = 1538$ \\ \hline
POD-Galerkin ROM $N_c=2$ (CPU) & 70 & $n = 1158, 751$ \\ \hline
POD-Galerkin ROM $N_c=4$ (CPU) & 33 & $n = 435, 365, 298, 45$ \\ \hline
\end{tabular}
\begin{flushleft} POD-Galerkin ROM and DL-ROM computational times along with FOM and reduced trial manifold(s) dimensions. DL-ROM provides a more efficient ROM with respect the POD-Galerkin ROMs.
\end{flushleft}
\label{tb:Tb3}
\end{table}

\begin{figure}[!ht]
\centering
\includegraphics[scale=0.23]{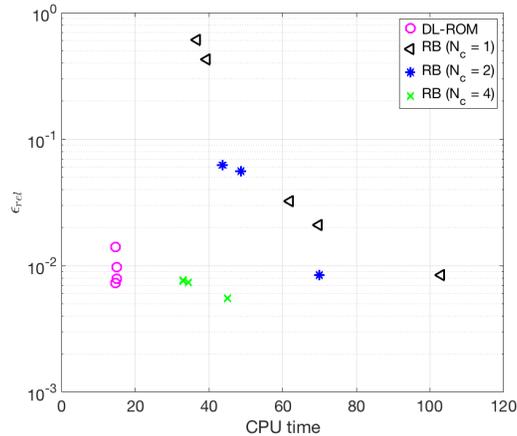}
\vspace{0.2cm}
\caption{{\bf Test 2: trend of the error indicator versus the CPU testing computational time.} Error indicator $\epsilon_{rel}$ vs. CPU testing computational time for different values of $N_c$ and $\varepsilon_{POD}$. The DL-ROM outperforms the POD-Galerkin ROM in terms of both efficiency and accuracy.}
\label{fig:Fig14}
\end{figure}

In \figurename~\ref{fig:Fig14} we show the trend of the error indicator (\ref{eq:error_indicator}) over the testing set versus the CPU  computational time both for the DL-ROM   and the POD-Galerkin ROM at testing phase. 
Slight improvements {of the performance of DL-ROM, in terms of accuracy,}  are obtained for a small increase of the DL-ROM dimension $n$, coherently with our previous findings reported in \cite{fresca2020comprehensive}. Indeed, the DL-ROM is able, also in this case, 
to accurately represent the solution manifold by a reduced nonlinear trial manifold of dimension $n_{\mu} + 1 = 2$; for the case at hand, we report the results for  $n = 5$ (very close to the intrinsic dimension $n_{\mu} + 1 = 2$ of the problem, and much smaller than the  POD-Galerkin ROM dimension), providing slightly smaller values of the error indicator (\ref{eq:error_indicator})  than in the case $n = 2$. Regarding instead the POD-Galerkin ROM, in \figurename~\ref{fig:Fig14} we report results obtained for different tolerances $\varepsilon_{POD} =$ 
$10^{-4}$, $5 \cdot 10^{-4}$, $10^{-3}$, $5 \cdot 10^{-3}$, $10^{-2}$.  In the cases  $N_c = 2$ and $N_c = 4$ we only report the results related to the smallest POD tolerances, which indeed allow us to meet the prescribed accuracy on the approximation of the gating variable, which would otherwise impact dramatically on the overall accuracy of the POD-Galerkin ROM. Moreover, we do not consider  more than $N_c = 4$ local reduced bases in order not to generate too small local linear subspaces. 
As shown in \figurename~\ref{fig:Fig14}, the proposed DL-ROM outperforms the POD-Galerkin ROM in terms of both efficiency and accuracy. 

\begin{figure}[!ht]
\centering
\includegraphics[scale=0.4215]{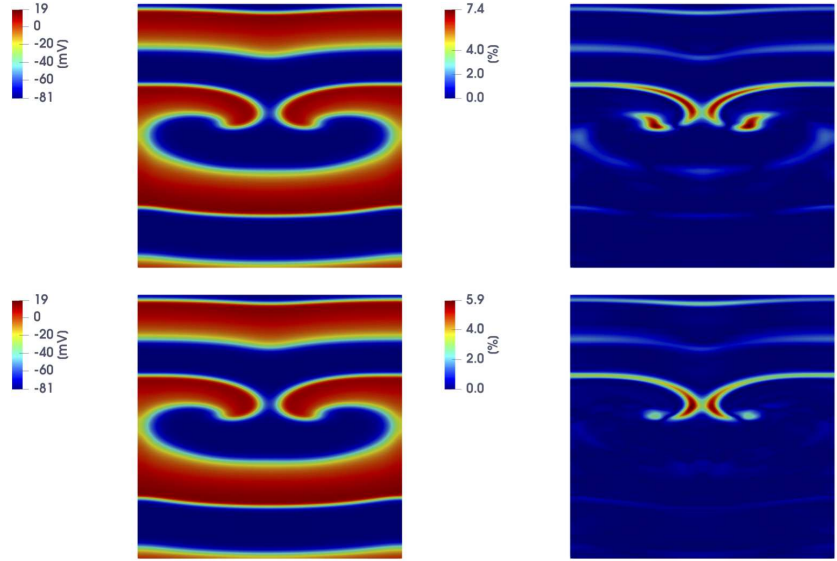}
\smallskip
\caption{{\bf Test 2: POD-Galerkin ROM solutions for different testing-parameter instances.} POD-Galerkin ROM solution (left) and relative error $\boldsymbol{\epsilon}_k^s$ (right) for $N_c = 2$ (top) and $N_c = 4$ (bottom) at $\tilde{t} = 157.2$ ms, for $\mu_{test} = 0.9625$ cm.}
\label{fig:Fig15}
\end{figure}

In \figurename~\ref{fig:Fig15} we show the solutions obtained through the POD-Galerkin ROM with  $N_c = 2$ (top) and $N_c= 4$ (bottom) local reduced bases, along with the relative error defined in (\ref{eq:relative_error_space}), for the testing-parameter instance $\mu_{test} = 0.9625$ cm at $\tilde{t} = 157.2$ ms. In both cases, we have considered the setting yielding the most efficient POD-Galerkin ROM approximation, which require about 30 (40, respectively) seconds to be evaluated. 
%
%
%
%
 %
By comparing \figurename~\ref{fig:Fig15} and \figurename~\ref{fig:Fig11} (bottom), we observe that the DL-ROM outperforms the POD-Galerkin ROM in terms of accuracy. 

\begin{figure}[!ht]
\centering
\includegraphics[scale=0.22]{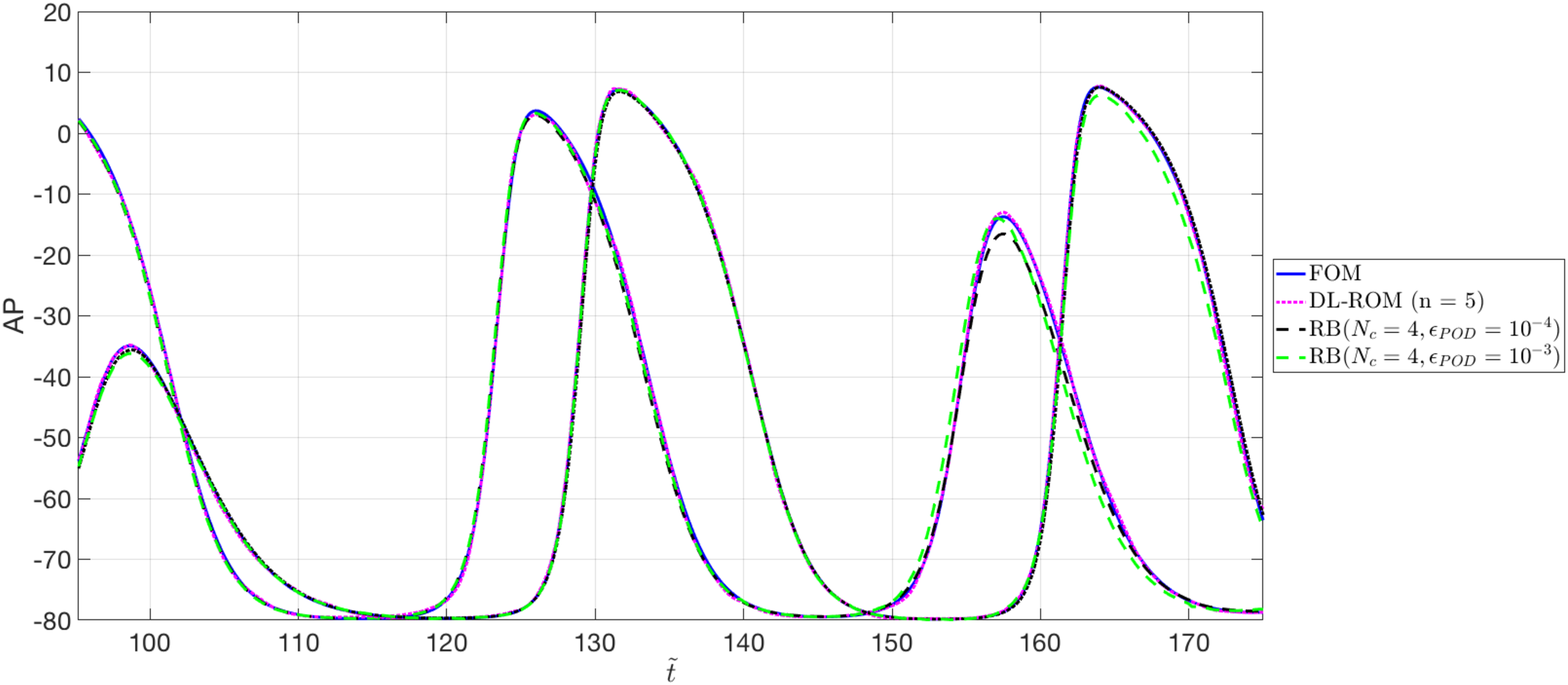}
\smallskip
\caption{{\bf Test 2: FOM, POD-Galerkin ROM and DL-ROM APs at $\mathbf{P_1}$ and $\mathbf{P_2}$.} AP obtained through the FOM, the DL-ROM and the POD-Galerkin ROM with $N_c = 4$,  for the testing-parameter instance $\mu_{test} = 0.9625$ cm, at $P_1 = (0.64, 1.11)$ cm and $P_2 = (0.69, 1.03)$ cm. The POD-Galerkin ROM approximations are obtained by imposing a POD tolerance  $\varepsilon_{POD} = 10^{-4}$ and $10^{-3}$, resulting in error indicator (\ref{eq:error_indicator}) values  equal to $5.5 \times 10^{-3}$ and $7.6 \times 10^{-3}$, respectively.}
\label{fig:Fig16}
\end{figure}

In \figurename~\ref{fig:Fig16} we show the action potentials obtained through the FOM, the DL-ROM and the POD-Galerkin ROM (with $N_c = 4$ local reduced bases), for the testing-parameter instance $\mu_{test} = 0.9625$ cm, and evaluated at  $P_1 = (0.64, 1.11)$ cm and $P_2 = (0.69, 1.03)$ cm. 
These two points are close to the left core of the figure of eight re-entry, where  a shorter action potential duration, and lower values of AP  due to the meandering of the cores, are observed. The AP dynamics at those points is accurately captured by the DL-ROM, while the  POD-Galerkin ROM might fail in this respect. 



\subsection*{Test 3: Three-dimensional ventricle geometry}

\begin{figure}[h!]
\centering
\includegraphics[scale=0.15]{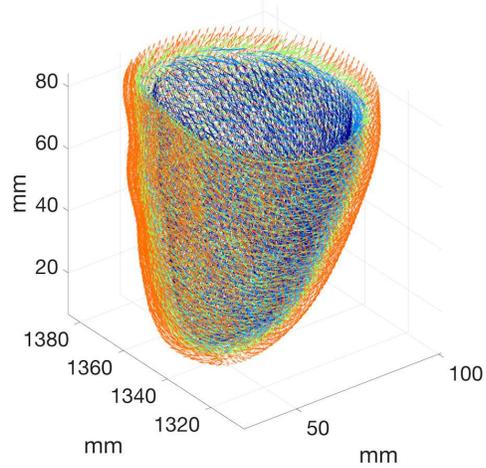}
\caption{{\bf Test 3: fibers distribution.}
Fibers field on the Zygote LV geometry.}
\label{fig:Fig17}
\end{figure}

We finally consider the solution of the coupled system (\ref{eq:monodomain}) in a three-dimensional left ventricle (LV) geometry, obtained from the 3D Human Heart Model provided by Zygote \cite{zygote2014}.
Here, we consider a single ($n_{\mu} = 1$) parameter, given by the longitudinal conductivity  in the fibers direction. The conductivity tensor takes the form 
\begin{equation}
\label{eq:D_ventricle}
\bf{D}(\mathbf{x}; \mu) = \sigma_t I + (\mu - \sigma_t) \mathbf{f}_0 \otimes \mathbf{f}_0,  
\end{equation}
where $\sigma_t = 12.9 \cdot 0.02$ mm\textsuperscript{2}/ms;   $\mathbf{f}_0$ is determined at each mesh point through a {\em rule-based} approach, by solving a suitable Laplace problem \cite{rossi2014thermodynamically}. The resulting fibers field is reported in \figurename~\ref{fig:Fig17}. 
The applied current is defined as   
\begin{equation*}
I_{app}(\mathbf{x}, t) = \frac{C}{(2 \pi)^{3/2} \alpha} \exp \bigg( - \frac{||\mathbf{x} - \mathbf{\bar{x}}||^2}{2 \beta} \bigg)\mathbf{1}_{[0, \bar{t}]}(\tilde{t}), 
\end{equation*}
where $\bar{t} = 2$ ms, $C = 1000$ mA, $\alpha = 50$, $\beta = 50$ mm\textsuperscript{2}, $\mathbf{\bar{x}} = {[44.02, 1349.61, 63.28]}^T$ mm. 

In order to build the snapshot matrix ${\bf S}$, we solve problem (\ref{eq:monodomain}) completed with the conductivity tensor (\ref{eq:D_ventricle}) by means of linear finite elements, on a mesh made by $N = 16365$ vertices, and a semi-implicit  scheme in time over a uniform partition of $(0,T)$  with  
$T = 300$ ms and time step $\Delta t = 0.1/12.9$. We uniformly sample $N_t = 1000$ time instances in  $(0, T)$ and we zero-padded \cite{goodfellow2016deep} the snapshot matrix to reshape each column in a 2D square matrix. The parameter space is provided by $\mathcal{P} = 12.9 \cdot [0.04, 0.4]$ mm\textsuperscript{2}/ms; here we consider $N_{train} = 25$ training-parameter instances and $N_{test} = 24$ testing-parameter instances computed as in Test 2. In this case, the maximum number of epochs is set to $N_{epochs} = 30000$, the batch size is $N_b = 40$ and 
 the training is stopped if the loss does not decrease over 4000 epochs.

\begin{figure}[h!t]
\centering
\includegraphics[scale=0.06]{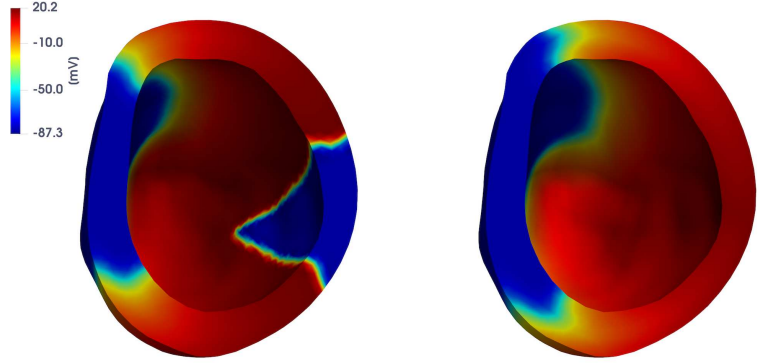}
\caption{{\bf Test 3: FOM solutions for different testing-parameter instances.}
FOM solutions for $\mu = 12.9 \cdot 0.0739$ mm\textsuperscript{2}/ms (left) and $\mu = 12.9 \cdot 0.1991$ mm\textsuperscript{2}/ms (right) at $\tilde{t} = 276$ ms. By increasing the value of $mu$ the wavefront propagates faster.}
\label{fig:Fig18}
\end{figure}

In \figurename~\ref{fig:Fig18} we report the FOM solution for two testing-parameter instances, i.e. $\mu = 12.9 \cdot 0.0739$ mm\textsuperscript{2}/ms and $\mu = 12.9 \cdot 0.1991$ mm\textsuperscript{2}/ms, at $\tilde{t} = 276$ ms, to show the variability of the FOM solution over the parameter space. As expected, front propagation is faster  for larger values of the parameter $\mu$. 

\begin{figure}[h!t]
\centering
\includegraphics[scale=0.175]{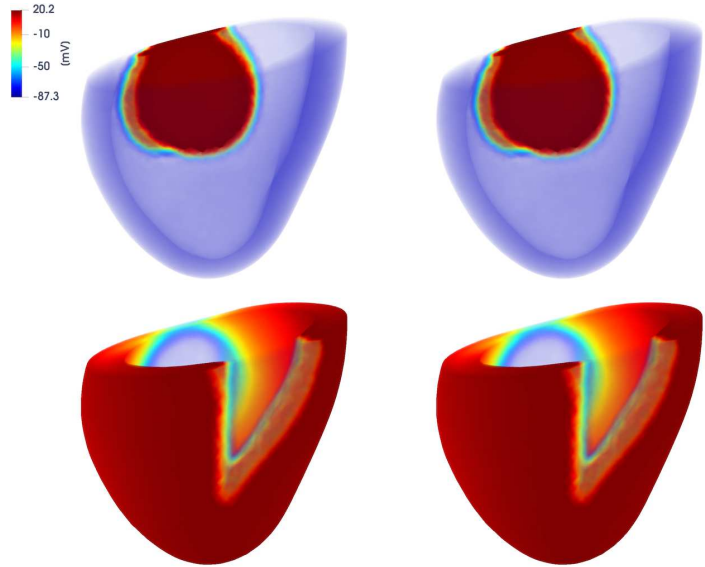}
\caption{{\bf Test 3: comparison between FOM and DL-ROM solutions for a testing-parameter instance at different time instances.}
FOM solution (left) and DL-ROM one (right), with $n = 10$, at $\tilde{t} = 42. 1$ ms (top) and $\tilde{t} = 276$ ms (bottom), for the testing-parameter instance $\mu_{test}= 12.9 \cdot 0.1435$ mm\textsuperscript{2}/ms.}
\label{fig:Fig19}
\end{figure}

\begin{figure}[h!t]
\centering
\includegraphics[scale=0.175]{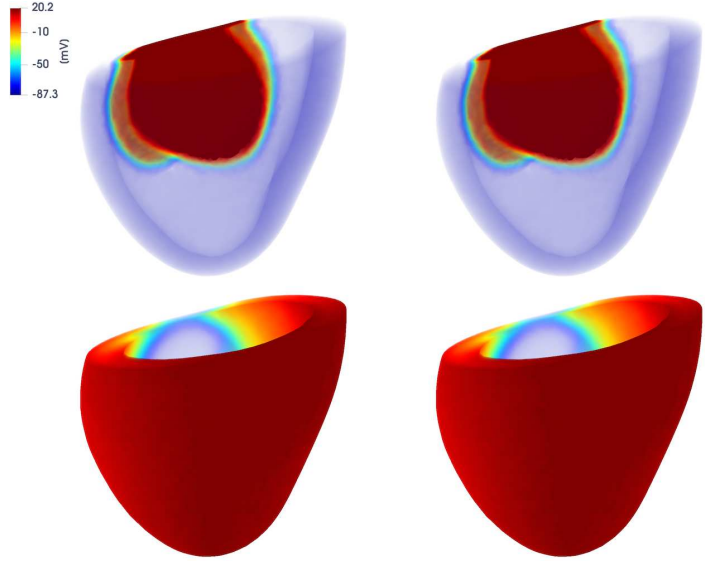}
\caption{{\bf Test 3: comparison between FOM and DL-ROM solutions for a testing-parameter instance at different time instances.}
FOM solution (left) and DL-ROM one (right), with $n = 10$, at $\tilde{t} = 42. 1$ ms (top) and $\tilde{t} = 276$ ms (bottom), for the testing-parameter instance $\mu_{test}= 12.9 \cdot 0.3243$ mm\textsuperscript{2}/ms.}
\label{fig:Fig20}
\end{figure}

In \figurename~\ref{fig:Fig19}-\ref{fig:Fig20} we report the FOM and DL-ROM solutions, the latter with $n = 10$, at $\tilde{t} = 42. 1$ ms and $\tilde{t} = 222.1$ ms, for two testing-parameter instances, $\mu_{test} = 12.9 \cdot 0.1435$ mm\textsuperscript{2}/ms and $\mu_{test} = 12.9 \cdot 0.3243$ mm\textsuperscript{2}/ms. The DL-ROM approximation is essentially as accurate as the FOM solution.  

 {Also for this test case, it is possible to build a reduced nonlinear trial manifold of dimension very close to the intrinsic one -- $n_{\mu} + 1 = 2$ -- as long as the maximum number of epochs $N_{epochs}$ is increased; the choice $n=10$ is obtained as the best trade-off between accuracy and efficiency of the DL-ROM approximation in this case.}

\begin{figure}[h!]
\centering
\includegraphics[scale=0.055]{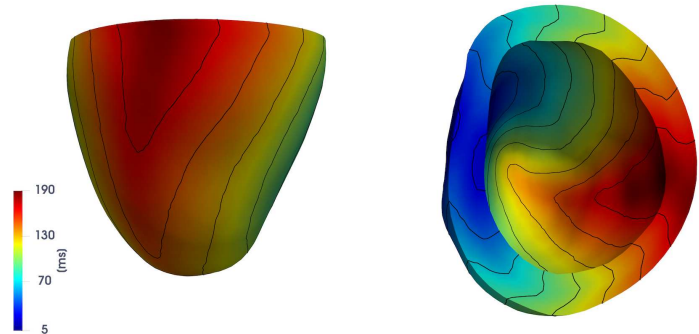}
\caption{{\bf Test 3: FOM activation map.}
FOM activation map for the testing-parameter instance $\mu_{test}= 12.9 \cdot 0.31$ mm\textsuperscript{2}/ms.}
\label{fig:Fig21}
\end{figure}

\begin{figure}[!ht]
\centering
\includegraphics[scale=0.055]{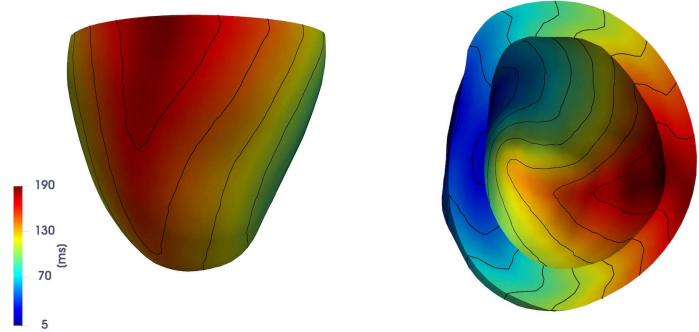}
\caption{{\bf Test 3. DL-ROM activation map.}
DL-ROM activation map for the testing-parameter instance $\mu_{test}= 12.9 \cdot 0.31$ mm\textsuperscript{2}/ms with $n = 20$.}
\label{fig:Fig22}
\end{figure}

{The DL-ROM approximation can also replace the FOM solution when evaluating outputs of interest. For instance, in \figurename~\ref{fig:Fig21} and \ref{fig:Fig22} we show the FOM and DL-ROM activation maps, the latter obtained by choosing $n = 10$ as DL-ROM dimension. Given the electric potential ${u}={u}({\bf x}, {t}; \boldsymbol{\mu})$, the (unipolar)  activation map at a point ${\bf x} \in \Omega$ is evaluated as the minimum time at which the AP peak reaches ${\bf x}$, that is,
\begin{equation*}
AC ( {\bf x}; \boldsymbol{\mu} ) = \arg \min_{{t} \in (0,T)}  \left ( {u}({\bf x}, {t}; \boldsymbol{\mu}) = \max_{{t} \in (0,T)} {u}({\bf x}, {t}; \boldsymbol{\mu})   \right). 
\end{equation*}
Here we compare the activation maps $AC_{FOM}$ and $AC_{DL-ROM}$  obtained through the FOM and the DL-ROM,  respectively, by evaluating the maximum  of the relative error  
\begin{equation*}
\boldsymbol{\epsilon}_{AC}({\bf x}; \boldsymbol{\mu})  = 
\frac{|AC_{FOM} ( {\bf x}; \boldsymbol{\mu} ) - AC_{DL-ROM} ( {\bf x}; \boldsymbol{\mu} )|}{|AC_{FOM} ( {\bf x}; \boldsymbol{\mu} )|}
\end{equation*}
over the $N$ mesh points; in the case $\mu = \mu_{test}= 12.9 \cdot 0.31$, the maximum relative error   is  equal to $4.32 \times 10^{-5}$.}

\begin{figure}[!ht]
\centering
\includegraphics[scale=0.41]{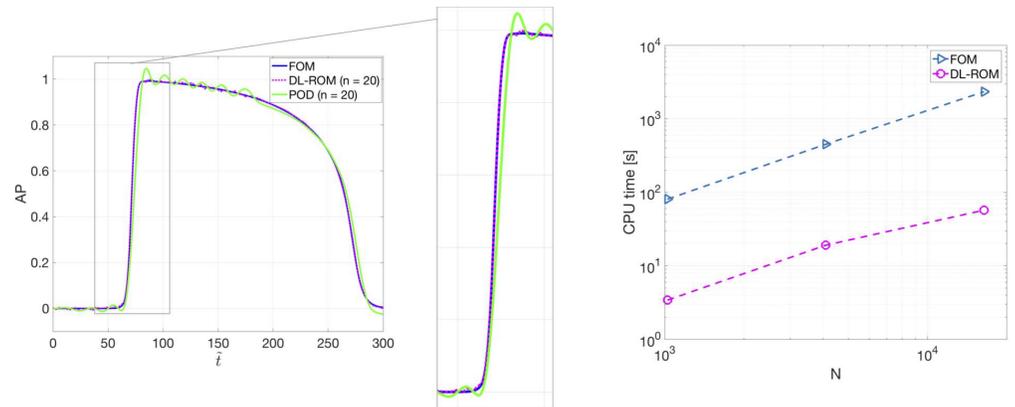}
\caption{{\bf Test 3: FOM, DL-ROM and optimal-POD APs for a testing-parameter instance. FOM and DL-ROM CPU computational times.}
FOM, DL-ROM and optimal-POD APs for the testing-parameter instance $\mu_{test}= 12.9 \cdot 0.31$ mm\textsuperscript{2}/ms (left). For the same $n$, the DL-ROM is able to provide more accurate results than the optimal-POD. CPU time required to solve the FOM and by DL-ROM at testing time with $n = 10$ vs $N$ (right). The DL-ROM is able to provide a speed-up equal to 41.}
\label{fig:AP_cputime}
\end{figure}

In \figurename~\ref{fig:AP_cputime} (left) we report the action potentials obtained with the FOM and the DL-ROM, this latter with $n = 20$, computed at point $P = [36.56, 1329.59, 28.82]$ mm for the testing-parameter instance $\mu_{test}= 12.9 \cdot 0.31$ mm\textsuperscript{2}/ms. Moreover, we also report the best approximation of the FOM action potential over a POD space of same dimension $n=20$ for the sake of comparison. Clearly, in dimension $n = 20$ the DL-ROM approximation is much more accurate than the POD best approximation; to reach the same accuracy (about $\epsilon_{rel} = 5.7 \times 10^{-3}$, measured through the error indicator (\ref{eq:error_indicator})) achieved by the DL-ROM with $n=20$,   $n = 120$ POD modes would be required.

In \figurename~\ref{fig:AP_cputime} (right) we highlight instead  the improvements, in terms of efficiency, enabled by the use of the DL-ROM technique. In particular, we point out the CPU time required to solve the FOM for a   testing parameter instance, and the one required by DL-ROM (of dimension  $n=10$) at testing time, {by using the time resolutioin each solution computation requires} and by varying the FOM dimension $N$ on a 6-core platform\footnote{Numerical tests have been performed on a MacBook Pro Intel Core i7 6-core with 16 GB RAM.} the FOM solution with  $N = 16365$ degrees of freedom requires about 40 minutes to be computed, against 57 seconds required by the DL-ROM approximation, thus implying a  speed-up almost equal to 41 times.

\section*{Conclusion}

In this work we have proposed a new efficient reduced order model obtained using deep learning algorithms to boost the solution of parametrized problems in cardiac electrophysiology. Numerical results show that  the resulting DL-ROM technique, formerly introduced in \cite{fresca2020comprehensive}, allows one to accurately capture complex wave propagation processes, both in  physiological and pathological scenarios.

The proposed DL-ROM technique provides ROMs that are orders of magnitude more efficient than the ones provided by common linear (projection-based) ROMs, built for instance through a POD-Galerkin reduced basis method, for a prescribed level of accuracy. Through the use of DL-ROM, it is possible to overcome the main computational bottlenecks shown by POD-Galerkin ROMs, when addressing parametrized problems in cardiac electrophysiology. The most critical points related to (i) the linear superimposition of modes which linear ROMs are based on; (ii) the need to account for the gating variables when solving the reduced dynamics, even if not required; and (iii) the necessity to use (very often, expensive) hyper-reduction techniques to deal with terms that depend nonlinearly on either the transmembrane potential or the input parameters, are all addressed by the DL-ROM technique, which finally yields more efficient and accurate approximation than POD-Galerkin ROMs.  {Moreover,  larger time resolutions can be employed when using a DL-ROM, compared to the ones required by the numerical solution of a dynamical systems through a FOM or a POD-Galerkin ROM. Indeed, the DL-ROM approximation can be queried at any desired time, without requiring to solve a dynamical system until that time, thus drastically decreasing the computational time required to compute the approximated solution at any given time.}

Outputs of clinical interest, such as activation maps and action potentials, can be more efficiently evaluated by the DL-ROM technique than by a FOM built through the finite element method, while maintaining a high level of accuracy. This work is a proof-of-concept of the DL-ROM technique ability to investigate intra- and inter- subjects variability, towards performing multi-scenario analyses in real time and, ultimately,  supporting decisions in clinical practice. In this respect, the use of DL-ROM techniques can foster  assimilation of clinical data with physics-driven computational models.

\section*{Supporting information}


\paragraph*{SI Code.}{\bf{Code and data.}} The code used in this work can be downloaded from: \url{https://github.com/stefaniafresca/DL-ROM}. The training and testing datasets will be made available upon request to the authors.

\paragraph*{SI Appendix.}
\label{SI_Appendix}
{\bf DL-ROM neural network architecture.} Here we report the configuration of the DL-ROM neural network used for our numerical tests. We employ a 12-layers DFNN equipped with 50 neurons per hidden layer and $n$ neurons in the output layer, where $n$ corresponds to the dimension of the reduced nonlinear {trial} manifold. The architectures of the encoder and decoder functions are instead  reported in \tablename~\ref{tb:table_transposed_convolutional_layers_encoder} and \ref{tb:table_transposed_convolutional_layers}.
\begin{table}[ht]
\centerline{\small
\begin{tabular}{|c|c|c|c|c|c|c|}
\hline
Layer & Input  & Output  & Kernel size & $\#$ of filters & Stride & Padding\\
 & dimension &   dimension &   & &  &  \\
\hline
1 & & & [5, 5] & 8 & 1 & SAME\\
\hline
2 & & & [5, 5] & 16 & 2 & SAME\\
\hline
3 & & & [5, 5] & 32 & 2 & SAME\\
\hline
4 & & &[5, 5] & 64 & 2 & SAME\\
\hline
5 & $N$ & 256 & & & & \\
\hline
6 & 256 & $n$ & & & &\\
\hline
\end{tabular}
}
\medskip
\caption{Attributes of convolutional and dense layers in the encoder $\mathbf{f}_n^E$.}
\label{tb:table_transposed_convolutional_layers_encoder}
\end{table}

\begin{table}[ht]
\centerline{\small
\begin{tabular}{|c|c|c|c|c|c|c|}
\hline
Layer & Input  & Output  & Kernel size & $\#$ of filters & Stride & Padding\\
  & dimension &   dimension &   & &  &  \\
\hline
1 & $n$ & 256 & & & &\\
\hline
2 & 256 & $N$ & & & &\\
\hline
3 & & & [5, 5] & 64 & 2 & SAME\\
\hline
4 & & & [5, 5] & 32 & 2 & SAME\\
\hline
5 & & & [5, 5] & 16 & 2 & SAME\\
\hline
6 & & & [5, 5] & 1 & 1 & SAME\\
\hline
\end{tabular}
}
\medskip
\caption{Attributes of dense  and transposed convolutional layers in the decoder $\mathbf{f}^D$.}
\label{tb:table_transposed_convolutional_layers}
\end{table} 

\section*{Acknowledgments}

The authors have been partially supported by the ERC Advanced Grant iHEART, ``An integrated heart model for the simulation of the cardiac function'', 2017-2022, P.I. A. Quarteroni (ERC2016AdG, project ID: 740132). Moreover, the authors acknowledge
{Dr. S. Pagani (MOX, Politecnico di Milano) for the kind help in the FOM software development, and Dr. M. Fedele for providing us the computational mesh of the Zygote Solid 3D heart model. 
\nolinenumbers

%
%
%


\end{document}